\documentclass[aps,prb,twocolumn,reprint,superscriptaddress,english]{revtex4-1}
\usepackage[utf8]{inputenc}
\usepackage[T2A,T1]{fontenc}
\usepackage[linktocpage=true,colorlinks=true,pdfborder={0 0 0},linkcolor=blue,citecolor=blue,filecolor=yellow,urlcolor=blue,bookmarks,pdfauthor={},]{hyperref}
\usepackage[full]{textcomp}
\usepackage[dvipsnames]{xcolor}
\usepackage{bm}
\usepackage{graphicx}
\usepackage{float}
\usepackage{wrapfig}
\usepackage{placeins}

\usepackage{amsmath}
\usepackage{gensymb}
\usepackage{nicefrac}
\usepackage{chemformula}
\usepackage{natbib}
\DeclareSymbolFont{cyrillic}{T2A}{cmr}{m}{n}
\DeclareMathSymbol{\Sha}{\mathalpha}{cyrillic}{216}
\usepackage{subfig}
\usepackage{braket}
\usepackage{dcolumn}
\usepackage{txfonts}
\usepackage[classicReIm]{kpfonts}
\usepackage{mathdots}
\usepackage{enumitem}
\usepackage{caption}
\usepackage{soul}
\usepackage{tabularx}

\newcommand{\LasVegas}{Department of Physics \&{} Astronomy, University of Nevada, Las Vegas, Las Vegas, Nevada 89154, USA}
\newcommand{\NEXCL}{Nevada Extreme Conditions Laboratory, University of Nevada, Las Vegas, Las Vegas, NV 89154, USA}

\begin{document}

\title{Machine learning using structural representations for discovery of high temperature superconductors}

\date{\today{}}

\author{Lazar Novakovic}
\affiliation{\LasVegas}
\affiliation{\NEXCL}
\author{Ashkan~Salamat}
\email{salamat@physics.unlv.edu}
\affiliation{\LasVegas}
\affiliation{\NEXCL}
\author{Keith~V.~Lawler}
\email{keith.lawler@unlv.edu}
\affiliation{\NEXCL}

\begin{abstract} 

The expansiveness of compositional phase space is too vast to fully search using current theoretical tools for many emergent problems in condensed matter physics. 
The reliance on a deep chemical understanding is one method to identify local minima of relevance to investigate further, minimizing sample space.
Utilizing machine learning methods can permit a deeper appreciation of correlations in higher order parameter space and be trained to behave as a predictive tool in the exploration of new materials.
We have applied this approach in our search for new high temperature superconductors by incorporating models which can differentiate structural polymorphisms, in a pressure landscape, a critical component for understanding high temperature superconductivity.
Our development of a representation for machine learning superconductivity with structural properties allows fast predictions of superconducting transition temperatures ($T_c$) providing a $r^2$ above 0.94.
\end{abstract}

\maketitle

\section{Introduction}

Modern exploratory syntheses require extensive assistance from a wide range of computational tools at every step. 
In the search for high temperature superconductors, a race has been roused with the discovery of hydrogen rich materials under high pressures exhibiting superconducting transitions approaching room temperature.\cite{Shipley2021,dCarbon}  
Approaches like crystal structure prediction (CSP) which evaluate the possible stable polymorphs and compositions of a system have been crucial in interpreting the results of solid-state synthesis efforts and compression experiments.\cite{LONIE2011372,AVERY2019274,LYAKHOV20131172,Calypso,Pickard_2011}
For superconducting systems, there is the further requirement beyond just stability of calculating the critical superconducting transition temperature ($T_c$) of a material which can be done ab initio with knowledge only of the material's lattice and atomic configuration.\cite{Nbqewmgsgp,sh2020,epw,PONCE2016116}
These approaches have been taken to identify almost all the known high-$T_c$ superconducting binary hydrides such as H$_3$S\cite{h3s}, LaH$_{10}$\cite{Drozdov2019, Kong2021, PhysRevLett.122.027001},  YH$_9$\cite{Kong2021}, CaH$_6$ \cite{Ma2022}, along with several others.  

Ab initio $T_c$ calculations can take on the order of days to months to complete depending on the atomistic complexity and the importance of anharmonicity to the material.\cite{SSCHA-Monacelli_2021}
Such calculations can be not only long, but also fragile to the assumptions necessary for good accuracy such as the density functional or screened Coulomb interaction.\cite{Novakovic2022,Shipley2020}
Likewise, recent observations of ternary and higher composition hydrides indicate a more tuneable pathway to achieving ambient conditions superconductivity,\cite{Snider2020,DiCataldo2021} but these more complex compositions beyond metal binaries demand significantly more computational efforts to computationally discover new materials.\cite{Shipley2020,Shipley2021,Sun2019LimgH,Liu2017,liu2017potential} 
Because  of these complications, there is a strong motivation to develop machine learning methods to rapidly and accurately predict a material's superconducting properties.
This has been aided by the decades of superconductivity research and experimental data including Supercon database\cite{supercon.nims.go.jp}, with over 30,000 superconductor measurements of $T_c$ and the corresponding chemical composition (ie stoichiometry).

Stoichiometric derived machine learning models for a superconductor's $T_c$ exist and use additional descriptors such as atomic mass, charge, number of atoms, and similar properties to yield good regression performance.\cite{chemM,Stanev2018} 
Yet, these models do not include any detail on the 3-dimensionality of the structure as would be needed for ab initio calculations, and thus fail to appreciate details such as the known evolution of the "superconducting dome" of $T_c$ as a function of pressure which arises from the structural changes of a material upon compression.
Consequently, difficulties arise when attempting to use data for materials which have a wide range of $T_c$s measured for several pressures when training these stoichiometric derived models.
For instance, recorded $T_c$s for specific compositions with high variances are often thrown out in these models.\cite{chemM,Stanev2018} 

Recently, a new machine learning model predicting superconducting $T_c$s which includes a description of the 3-dimensional atomic structure has been developed.
This model utilizes the smooth overlap of atomic positions (SOAP)\cite{soap1,soap2,soap3,doi:10.1021/acs.chemrev.1c00021,Townsend2020} descriptors,  which incorporates information about the local distributions of positions of atoms around each atom within the structure and has shown improved performance compared to composition models.\cite{soap.model} 
While the incorporation of the SOAP descriptors provides much needed structural information, the description of local atomic distributions with a cutoff radius could overlook potential similarities between polymorphs within certain values of the cutoff radius as well as neglecting any properties that arise from the full periodic structure of the material such as the phonon dispersion.\cite{ALLEN19831}

Establishing descriptors capable of accurately capturing structure such as the atomic positions and lattice which affect superconducting properties is thus of tantamount importance for advancing these machine learning models. 
Arrangement of periodic atoms should be quantified in a normalized manner for all crystal structures while being weakly dependent on physically invariant transformations such as swapping labels of atoms of identical species or supercells. 
Here we develop a representation of structural properties for use with machine learning models which boosts predictive precision and extends capabilities for discovery.
This representation differentiates polymorphisms and provides physical interpretability of the structural properties.
These new descriptors go beyond the SOAP model's distance based description of the local structure of the material and incorporates the more nuanced properties of the periodic mass and charge distribution of the material.
Using the model's fast prediction speed, we study change in $T_c$ with lattice and atomic species variations of predicted superconducting materials.  

\section{Data}

\subsection{Supercon} 
The supercon database contains several thousands of various reported superconducting critical temperatures $T_c$ along with the superconductor's chemical compositions and the respective experimental paper(s).\cite{supercon.nims.go.jp} 
This data needed to be cleaned due to ambiguities and errors, and we chose to use the cleaned version of this data that used in a previous chemical composition machine learning model.\cite{chemM}
The data was reformatted by first grouping the sets of $T_c$s for distinct compositions as is illustrated in Table \ref{tbl:datatables}.
Averages and variances of the $T_c$s for a given composition were taken to compare the similarity of measured $T_c$s. 
Compositions with multiple reported temperatures were delineated into two sets by a cutoff, $\sigma^2 _{T_{cut}}$. 
This demarcation helps determine when temperature averages can accurately represent each set of structures; choosing the cutoff $\sigma^2 _{T_{cut}} = 2$ results in the set $\sigma^2_{T(y)} < \sigma^2 _{T_{cut}}$ having averaged measured $T_c$s within a few Kelvin.
Here, $y$ and $T(y)$ represent the composition and $T_c$ measurements respectively.
    
\begin{table}
\begin{tabular}{|l|l|l|l|}
\hline
Superconductor & $T_c$s (K)& $\braket{T_c}$ (K)& $\sigma^2$ (K$^2$)\\
\hline
BaLa$_9$(CuO$_4$)$_5$ &	29.0, 28.0, ... &	24.4 & 	57.1\\
BaLa$_{19}$Cu$_9$AgO$_{40}$ &	26.0, 27.0 &	26.5 &	0.25\\
BaLa$_{19}$(CuO$_4$)$_{10}$ &	19.0, 26.9,  ... &	28.2 &	31.9\\
Ba$_3$La$_{37}$(CuO$_4$)$_{20}$ &	22.0, 30.0, ... &	27.0 &	94.3\\
Ba$_3$La$_{17}$(CuO$_4$)$_{10}$	& 23.0, 9.7, ...	& 16.2 & 29.5\\
$\vdots$	& $\vdots$ &	$\vdots$ & $\vdots$ \\
CeBiS$_2$O &	1.6, 3.0 &	2.3 &	0.49\\
TiSiIr &	1.42, 1.85, ... &	2.23 &	0.75\\
Tm$_{21}$Lu$_4$(Fe$_3$Si$_5$)$_{25}$ &	2.44 &	2.44&	0.00\\
Nb$_4$Pd &	1.98 &	1.98&	0.00\\
Nb$_{69}$Pd$_{31}$ &	1.84 &	1.84&	0.00\\
\hline	
\end{tabular}
\caption{Compositions and $T_c$s for superconductors of the supercon dataset. 
$\braket{T_c}$ denotes the average of $T_c$s for the superconductor and $\sigma^2$ is the variance of the compositions $T_c$s.}
\label{tbl:datatables}
\end{table}
    
\subsection{Materials Project}
        
To add structural information to the $T_c$ values harvested from the Supercon dataset, the compositions from the Supercon data are matched with the corresponding structures on the Materials Project database\cite{matp,Ong_2015} with the pymatgen library.\cite{pymatgen1} 
We found 2454 Materials Project structures corresponding with the Supercon compositions, denoted by $S(y)$.
For the compositions with a low variance, $\sigma^2_{T(y)}<\sigma^2 _{T_{cut}}$, the gathered $T_c$ values were averaged and that $T_c$ was assigned to the corresponding Materials Project structure.
The high variance compositions with $\sigma_{T(y)}^2 \ge \sigma^2 _{T_{cut}}$ (which represents less than 10\% of the data could be treated with at least two different approaches.

Firstly, this data could be excluded because of the impreciseness with setting the mean to structures which allows for more consistent model accuracy. 
Alternatively, the structures can be labelled with measurements in the respective $T_c$ set that produce the model with the best performance on the validation set. Using a Bayesian process to determine the labelling of high variance data would be helpful to achieve this.
Although both of these approaches lead to models with better performance on the validation set, averaging this data is the method with the least data fitting.
For this analysis, we chose the third option of averaging this data to study baseline performance. 
        
\begin{figure}[h!]
    \centering
    \includegraphics[width=0.4587\textwidth]{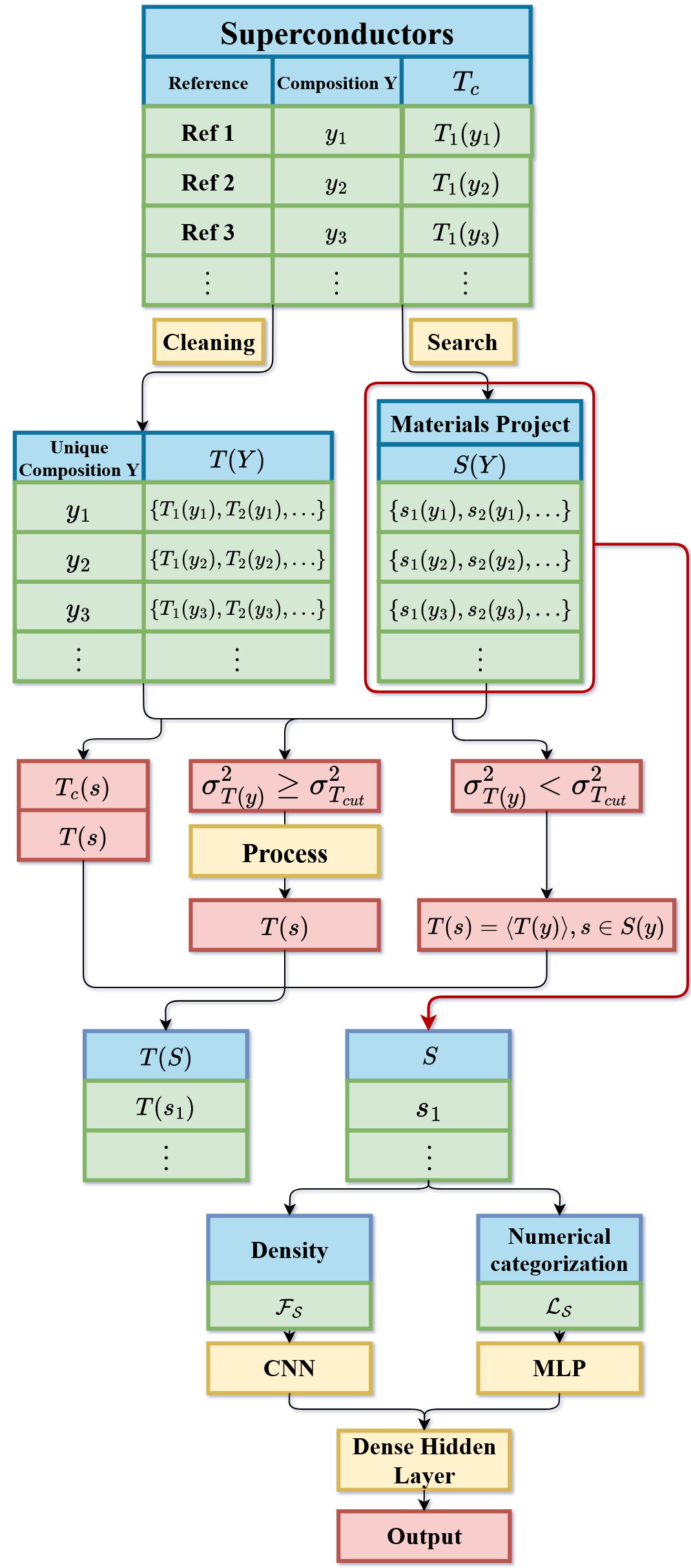}
    \caption{Data flow for the supercon data and materials project structures. High variance data can be processed separately from the averaged low variance sets.  Structures are described by densities of atomic properties and numerical descriptors of the lattice and composition. Together these are inputs of the convolutional neural network (CNN) and multilayer perceptron (MLP) ensemble.}
    \label{fig:data}
\end{figure}

\subsection{Theoretical Structures}

As one of the current motivations for machine learned $T_c$ models is the \textit{a priori} prediction of high pressure hydrogen-based superconductors, theoretical materials that fall into this category but are not within the Supercon dataset were included in our dataset.
As these materials are found at higher pressures, there is an associated superconducting dome wherein the $T_c$ varies with pressure.
Related, there is also a large degree of potentially accessible compositions and polymorphs within the pressure ranges that are studied experimentally for these materials.
46 high temperature theoretical hydride superconductors of various phases have been introduced into the data set for modelling.\cite{Shipley2021} 
The $T_c$s are labelled by the average of the $\mu^*$ bounds used when the $T_c$ was computed through directly solving the Eliashberg equations. 

\section{Representation}

\subsection{Fourier}
To establish descriptors, structures need first to be represented consistently and transformed to a descriptive set of numbers.
Atoms, $\mathbb{A}$, are described by which element they are $E\in \mathbb{E}= \{\text{H, He, Li, \dots}\}$ along with a position in space, $\mathbb{A} =\mathbb{E}\times\mathbb{R}^3$.
Following crystallographic norms, the structures in the data set are described by any collection of atoms and lattice vectors expressed as $\mathcal{P}(\mathbb{A})\times\mathbb{V}$. 
The set $\mathbb{V} \subseteq GL_3(\mathbb{R})$ are the lattice vectors with elements

\begin{equation}
\mathcal{V} = \vec{v}_i = 
\begin{pmatrix} 
        \vec{a} \\
        \vec{b} \\
        \vec{c}
\end{pmatrix}.
\end{equation}

To transform this structural data for the model in a normalized way, atoms are described as fields of their atomic mass and charge by their atomic number as they are among an atom's the most identifying characteristics. Field values on a discretized space provides a 3 dimensional grid input compatible with convolutional techniques. 
Studying distributions about atoms such as with SOAP \cite{soap1,soap2,soap3} does not provide global structural information or the distribution of mass, which are key interests for understanding superconductors in theory.

Only representing values at the atom's position results with a mostly zero-valued field. This also leads to difficult comparison between structures as small perturbations cause immediate drops and spikes in the values. Distributing the atoms as a density allows a nuanced description of the atomic properties as local regions of charge or mass akin to the description of a material that naturally arises from density functional approaches. Perturbations will result as gradual transitions in the overall density. 
        
The Gaussian density for a single atom of elemental species Z at position $\vec{R}_0$ in crystal coordinates is taken to be 
\begin{equation}
    \psi_{Z,\Gamma} \left(\vec{R};\vec{R_0}\right) = \frac{N_{Z,\Gamma}}{({\sqrt{2 \pi } \sigma_{Z,\Gamma} )^3}}
    e^{- \frac{\left (\vec{R}-\vec{R}_{0}\right)^2}{{2 \sigma^2 _{Z,\Gamma}}}}.
    \label{eq:psiunit}
\end{equation}
The atomic density for species Z is normalized to atomic charge and mass, indexed by $\Gamma$.
\begin{equation} N_{Z,\Gamma} = \{e(Z), m(Z)\}\end{equation}      
The Gaussian widths are taken to be physically relevant values. The mass width is proportional to the atomic number and the charge is taken to be proportional to the outer atomic radii. Each is scaled with parameters $\gamma$ to ensure satisfactory overlap.
        \begin{equation}
        \sigma^2 _{Z,\Gamma} = \{ \gamma_e e(Z), \gamma_m r(Z)\}
        \end{equation}
Consequently, this also encodes physical characteristics of the atoms in the field representation. Mass and outer atomic radii are key quantities for prediction in other composition based machine learned superconductivity models.\cite{Stanev2018}

        Introducing cell periodicity makes the density independent of unit cell choice, however expands the domain beyond a finite region. 
        The fully periodic density is the sum over all cells
\begin{equation}\Psi_{Z,\Gamma}(\vec{R};\vec{R}_0)= \sum_{tuv=-\infty} ^{\infty} \frac{N_{Z,\Gamma}}{ (\sqrt{2 \pi } \sigma_{Z,\Gamma}) ^3} e^{- \frac{(\vec{R}-\vec{R}_0 -\vec{\Delta}_{tuv})^2}{2 \sigma^2 _{Z,\Gamma} }} \\ 
\label{eq:fieldperiodic1}
\end{equation}

        \begin{equation}{\vec{\Delta}_{tuv}=t a\hat{a} + u b \hat{b} + v c\hat{c} }.
        \end{equation}
This is just a convolution of Dirac comb (Eq \ref{eq:DiracCom}) and the unit cell density (Eq \ref{eq:psiunit})
\begin{equation}
        \Sha_{\vec{\Delta}} = \sum_{tuv = -\infty}^{\infty} \delta^3 (\vec{R}-
        \vec{\Delta}_{tuv})\label{eq:DiracCom}\end{equation}
        
        \begin{equation}\Psi_{Z,\Gamma}(\vec{R};\vec{R_0}) = \left(\psi_{Z,\Gamma} (\vec{R_0}) \ast \Sha_{\vec{\Delta}} \right )\left(\vec{R}\right). \label{eq:fieldperiodic2}\end{equation}
With the periodic field density, the field may be expressed succinctly with far fewer terms through coefficients in

 \begin{equation}
         \Psi_{Z,\Gamma}(\vec{R};\vec{R}_0)= \sum_{hkl} \mathcal{F}_{Z,\Gamma}(\vec{h})e^{  i\vec{\bar{h}}\cdot\vec{R}}.
        \end{equation}

Using shorthand $\vec{h}$ with components h, k, and l, and  $\vec{\bar{h}}$ with the components ${\vec{\bar{h}}_i = 2\pi \vec{h}_i } / {|\vec{v}_i|}.$ 
Eqs \ref{eq:fieldperiodic1} and \ref{eq:fieldperiodic2} permit obtaining $\mathcal{F}$ by the Poisson summation formula as 
 \begin{equation}
            \mathcal{F}_{Z,\Gamma}({\vec{h}};\vec{R}_0)=\frac{1}{V}F\{\psi_{Z,\Gamma}\}\left(\frac{h}{a},\frac{k}{b},\frac{l}{c};\vec{R}_0\right).\end{equation}
Using cartesian coordinates, with lattice vectors $\mathcal{V}$, the analogous expression yields
        \begin{equation} \mathcal{F}_{Z,\Gamma} = \frac{1}{V}F\{\psi_{Z,\Gamma}\}(\mathcal{V}^{-1}\cdot \vec{h}).\end{equation}
        
         $F\{\psi_{Z,\Gamma}\}$ \normalsize being the Fourier transform of the original density and h,k,l as integer values

        \begin{equation}
        \begin{split}
        F\{\psi_{Z,\Gamma}\}& =  \iiint_{\mathbb{R}^3} d^3R \psi_{Z,\Gamma}(\vec{R};\vec{R}_0) e^{-2\pi i \vec{h}\cdot \vec{R}} \\ 
                            & =  N_{Z,\Gamma} e^{-2\pi^2 \sigma^2 _{Z,\Gamma} \vec{h}^2} e^{-2 \pi i \vec{h}\cdot\vec{R}_0}.
        \end{split}
        \end{equation}
        
        Coefficients are therefore given by 
        \begin{equation}\mathcal{F}_{Z,\Gamma} (\vec{h};\vec{R}_0) = \frac{N_{Z,\Gamma}}{V}e^{-\frac{1}{2}{\sigma_{Z,\Gamma}^2 \vec{\bar{h}}^2}} e^{-i\vec{\bar{h}}\cdot \vec{R_0}}.
        \label{eq:coefficients}\end{equation}

The field can similarly have a periodic localized representation by the von Mises distribution which has periodicity built in and is similar in form to a Gaussian.
        
        \begin{equation}\Psi^{vm} _{Z,\Gamma}(\vec{R};\vec{R}_0)= \frac{N_{Z,\Gamma}}{ I_0 ^3 (\kappa_{Z,\Gamma})} \prod_j e^{\kappa_{Z,\Gamma} cos\left(\frac{2 \pi }{|\vec{v}_j|}(R_j-R_{0j} )\right)} \end{equation}

$I_0(\kappa_Z)$ \normalsize being the modified Bessel function of the first kind and  $\kappa_Z \propto \frac{1}{\sigma^2 _Z}$. \normalsize
        
        The Fourier coefficients are given by the modified Bessel functions of the first kind.
        
        \begin{equation} \mathcal{F}^{vm} _Z (\vec{h};\vec{R}_0) = \frac{N_Z}{V I_0 ^3 (\kappa_Z)} I_{\bar{h}}(\kappa_Z) I_{\bar{k}}(\kappa_Z) I_{\bar{l}}(\kappa_Z)  e^{-i \vec{\bar{h}} \cdot \vec{R}_0} \end{equation}
        $ I_{\bar{h}}(\kappa_{Z}) $  increases with $\kappa_{Z}$, meaning the ratios will tend to be of large values. 
        In the region of $\kappa_Z >> \bar{h} $ this ratio goes to
        \begin{equation}
        \frac{I_{\bar{h}} (\kappa_Z)}{I_0(\kappa_Z)}  \rightarrow e^{-\frac{{1}}{2\kappa_Z}\bar{h}^2} \end{equation}
        This shows the similar shape and values of the densities, as expected.

The resulting fields make the input for the spatial data.
    \begin{equation}\mathcal{F}_Z =\mathcal{F}_{Z,e(Z)} \otimes \mathcal{F}_{Z,m(Z)}\end{equation}

The coefficients for the entire structure are then the sum of the each atomic contribution of $\mathcal{F}$ by linearity, and the inverse uniquely recreates the density field.

This representation transforms positions of atoms in a lattice to two 3d grids of complex coefficients.

\begin{equation} \mathbb{F}_{\gamma} : \mathbb{A}\times \mathbb{V} \rightarrow \left ( \mathbb{C} \otimes \mathbb{C}\right )^{ \mathbb{Z}^3}\end{equation}

        An atom in a lattice, $\mathcal{\Tilde{A}}\in\mathbb{A}\times \mathbb{V} $, represents each parameter needed in generating the coefficients of the densities, $ Z=\pi_1 (\mathcal{\Tilde{A}}) $, $\vec{R}_0 = \vec{\pi}_2 (\mathcal{\Tilde{A}})$, $\mathcal{V} ={\pi}_{3} (\mathcal{\Tilde{A}})$.
        \begin{equation} \ \mathbb{F}_{\gamma} (\Tilde{\mathcal{A}}) = \mathcal{F}_{Z}(\vec{h};\vec{R}_0)\end{equation}

        The Fourier coefficients of the total structure $\mathcal{S}$ is by summing for each atom.
        
        \begin{equation}\mathcal{F_S} = \sum _{\Tilde{\mathcal{A}}\in \mathcal{S}} \mathbb{F_{\gamma}(\Tilde{\mathcal{A}})}\end{equation}
Some coefficients are redundant as equation \ref{eq:coefficients} is symmetric for many points and is hermitian. Taking some subset of the coefficients approximates and reduces the data significantly.
        Reducing by cutting off the coefficients to $0\leq h,k,l < \delta$ yields a $\delta\times\delta\times\delta\times 2$ set of spatial data to represent the structure. 

\subsection{Categorical and Numerical}
    
    There are more parameters which are useful to characterize the superconductors beyond the structural properties of atomic densities in a unit cell.
    In total 175 numerical descriptors were used for this data.
    Categorical and numerical data consisted of the number of each species in the unit cell up to N = 86, 81 chemical descriptors utilized in previous composition model\cite{chemM}, along with the 8 descriptors with the lattice parameters and angles, space group number, and volume found using pymatgen.\cite{pymatgen1}

\section{Model}
\subsection{Input and Parameters}
The processing parameters necessary for the structural ensemble are: cutoff $\delta$, scaling factors $\gamma_e$ and $\gamma_m$, and $\mathcal{F}$ type and domain.

The form of the descriptors is 3d spatial data having 2 channels which can be modelled through a convolutional neural network (CNN).\cite{LeCun2015DeepL} 
Conventional neural networks are real-valued, however the output of the density is generally complex meaning the input for the convolutional layer will be complex.
A complex-valued convolutional neural network (CVCNN) may be appropriate for properly modelling. The data may be made to be real-valued by taking the modulus, real, or imaginary components. 
This allows use of standard 3d convolutional neural networks although decreases the information of the structure.
The data can be treated with a single channel with all the complex data included or with separate channels for each real and imaginary part keep that retain the complete information at a cost of double the data size and connection between the terms. 
Each of these seemed to work properly relative to each other, though a CVCNN with the cvnn library \cite{cvnn} showed the best performance generally. 

Various choice of model parameters were used and checked. For this work we chose $\delta = 13$, $\gamma_{e,m} = \{ 0.0025,0.25\}$, and $0\leq$ h,k,l for the processing. Optimal values can be finely tuned, though many configurations, including the chosen parameters, successfully yielded accurate models with novel behavior. The output was taken to be the exact complex values therefore a complex neural network was used.
This corresponded to 13$\times$13$\times$13$\times$2 complex data in the convolution port of the network.

\subsection{Architecture}

\begin{figure*}[t]
\centering
\includegraphics[width=\textwidth]{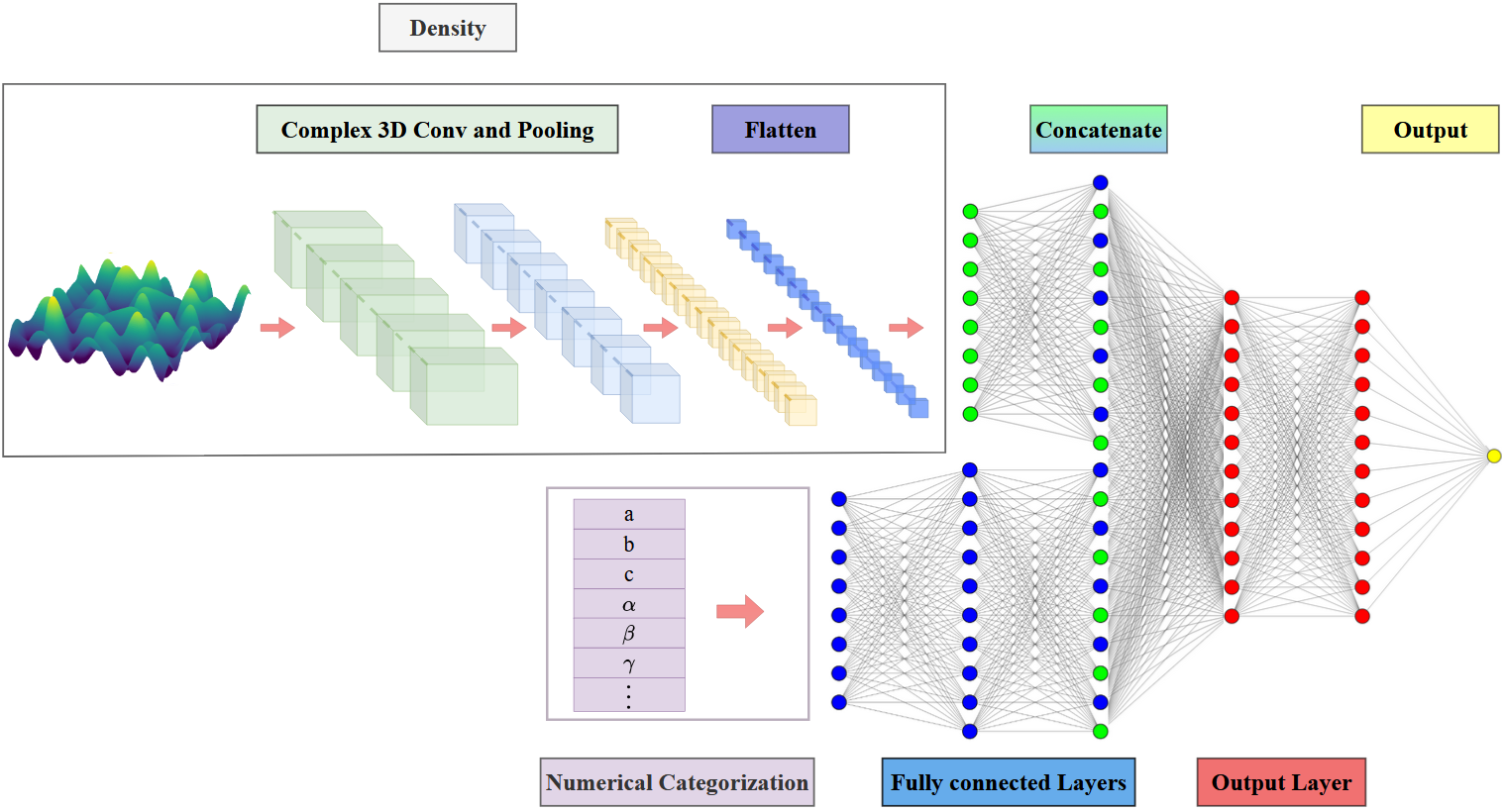}
\caption{The architecture of the model. The density information goes through a complex convolution process then flattened. Categorical and numerical data goes to a fully connected layer and is concatenated with the density output.}
\label{fig:ark}
\end{figure*}

The model comprises of two input layers with a convolutional layer for spatial density data and a dense neuron layer for the numerical data expressed by figure \ref{fig:ark}. 
TensorFlow  is used as well as the cvnn package \cite{cvnn} for the convolutional layers for building model. Keras tuner was used to tune the number of layers, neurons per layer, and loss function.\cite{tensorflow2015-whitepaper,omalley2019kerastuner} 
The numerical data is 4 fully connected 256 unit layers. Spatial data is taken through 2 convolutional layers consisting of 32 units of $3\times3\times3$ kernels.
$2\times2\times2$ average pooling and 50 percent dropout is used between the layers. A cartesian ReLu activation function is used for the first layer.\cite{tensorflow2015-whitepaper} 
The second layer uses an absolute value activation function to convert to real-valued output. 
This is flattened and then concatenated with the numerical data output layer which follow to 4 fully connected ReLu layers with 512 units and 50 \% dropout and a final single neuron output matching output $T_c$.\cite{tensorflow2015-whitepaper}

In training, a train-test split of 80-20 was used. 
Many loss functions have been tested in order to best fit the data which includes both several $T_c$s less than 1 K and above 100 K.
The keras Huber loss function with delta=7 was used to prevent outlier data from dominating the error.\cite{tensorflow2015-whitepaper} 
The best validation loss model is picked from 300 epochs of training.

\section{Results}
    \subsection{Validation}

\begin{figure}[h!]

            \includegraphics[width=0.42\textwidth]{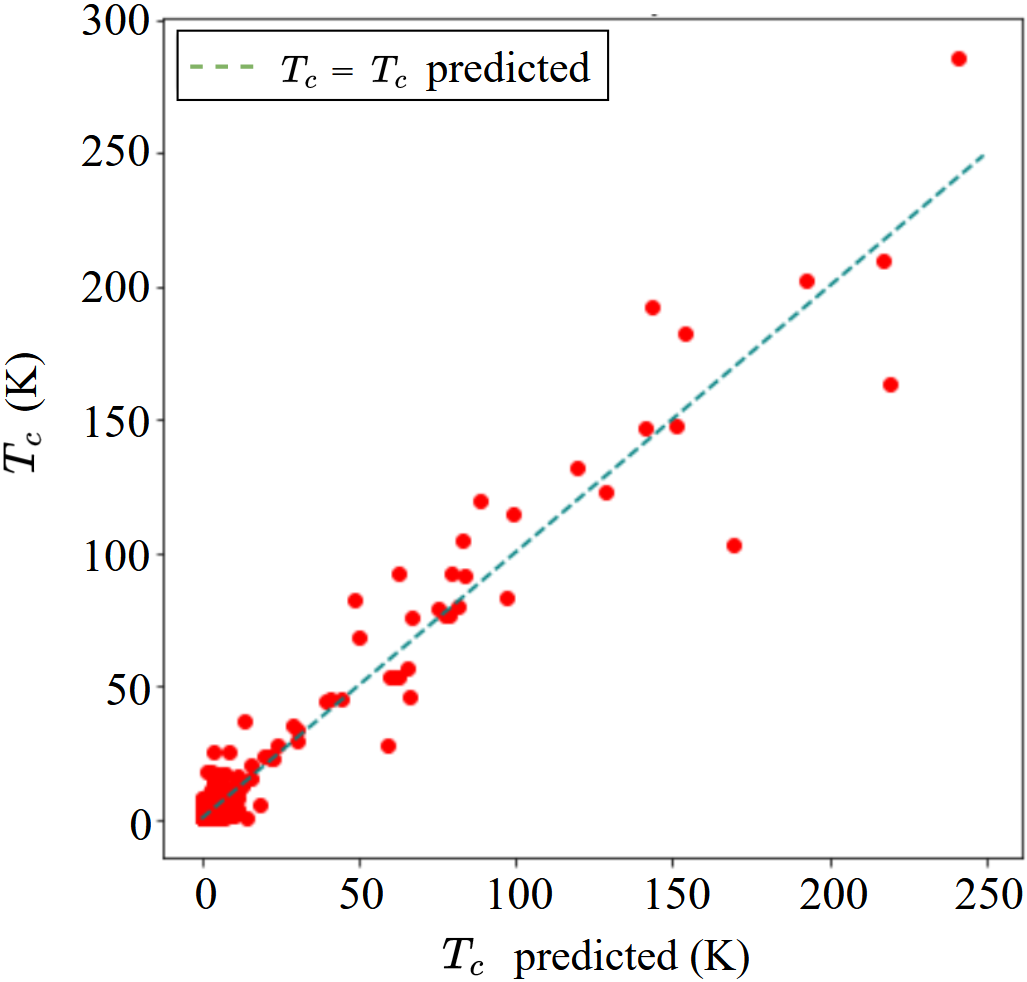}
            \caption{$T_c$ labelled vs. $T_c$ predicted by the model in the testing set. Where the values equal is expressed by the dashed green line. The $r^2$ is 0.9429. }
            \label{fig:Tc}
    \end{figure}

   The model yielded a $r^2 =$ 0.9429 with the validation set which shows great performance over the entire prediction space expressed in figure \ref{fig:Tc}. This is without exclusion of any data with high $T_c$ variances. Exclusion of these along with optimizing model architecture and processing of the structure would likely boost the performance as well minimize the number of outliers. 
    Increasing data on $T_c$s of polymorphisms should also refine the model accuracy, considering the potential number of missing structures which are phases of the superconductors used in the data.

 \begin{table}[t]
     \begin{tabular}{|l|l|l|l|}
        \hline
         Superconductor &$T_c$ (K) pred&$T_c$ (K) chem&$T_c$ (K)\\
       
        \hline
          LaH$_{10}$ C2/m (250 GPa)&  217   &22&215\cite{Shipley2020} \\
          YH$_{10}$ R$\bar{3}$m (400 GPa)&  {268 }  &       {16 }          &265\cite{Shipley2020}\\
          LaH$_{10}$ Fm$\bar{3}$m (250 GPa)    &  {241 }  &       {21 }          &246\cite{Shipley2020}\\
          YH$_{10}$ Fm$\bar{3}$m (400 GPa)     &  {263 }  &       {16 }          &265\cite{Shipley2020}\\
      
          LaH$_{10}$ R$\bar{3}$m (250 GPa)                &  {236 }  &       {21 }          &245\cite{Shipley2020}\\
  
        \hline
         SeH$_{3}$ Im$\bar{3}$m (200 GPa)  & 111 & 28 & 110\cite{seh3}\\
          CSH$_5$ Cm (150 GPa)                    &  {109 } &       {37 }          &103\cite{Cui2020}\\
          CSH$_7$ Pnma (200 GPa)                  &  {105 } &       {36 }          &129\cite{Cui2020}\\
              CSH$_7$ R3m (270 GPa)         &  {177   } &       {36 }          &174\cite{Novakovic2022}$^{*}$\\
         CSH$_7$ I43m (270 GPa)        &  {162    } &       {36 }          &159\cite{Novakovic2022}$^{*}$\\
        CSH$_7$ CmCm (200 GPa)                  &  {123 } &       {36 }          &-\cite{Cui2020}\\
          CSH$_7$ P$\bar{4}$m2 (200 GPa)                 &  {155 } &       {36 }          &-\cite{Cui2020}\\
          CSH$_7$ P3m1 (200 GPa)                  &  {188 } &       {36 }          &-\cite{Cui2020}\\
          CSH$_7$ P$\bar{4}$3m (200 GPa)                 &  {191 } &       {36 }          &-\cite{Cui2020}\\
        \hline
         LaH$_{10}$ Fm$\bar{3}$m (150 GPa)    &  {235 } &       {36 }          &249\cite{Drozdov2019}\\
    
         YH$_4$ I4mmm (167 GPa)        &  {44 } &       {21 }          & 82\cite{Wang2022}\\
         YH$_6$ Im$\bar{3}$m (201 GPa)        &  {174 } &       {25 }          &211\cite{Kong2021} \\
   
        YH$_9$ P6$_3$/mmc (255 GPa)         &  237 &       14          &237\cite{Kong2021}\\ 
    
        \hline

        Li$_2$MgH$_{16}$ {P$\bar{3}$m1 (300 GPa)}         &  {277 } &      {30 }          &191\cite{Sun2019LimgH}\\
        Li$_2$MgH$_{16}$ {Fd$\bar{3}$m (250 GPa)}         &  {293 } &      {30 }          &452\cite{Sun2019LimgH}\\
        Li$_2$MgH$_{16}$ {Fd$\bar{3}$m (300 GPa)}         &  {295 } &      {30 }          &315\cite{Sun2019LimgH}\\
        Li$_2$MgH$_{16}$ P1 (300 GPa)           &  {185 } &      {30 }          &-\cite{Sun2019LimgH}\\
        Li$_2$MgH$_{16} \alpha$-P-1 (300 GPa)   &  {154 } &      {30 }          &-\cite{Sun2019LimgH}\\
        Li$_2$MgH$_{16}$ Pm (300 GPa)            &  {183 } &      {30 }          &-\cite{Sun2019LimgH}\\
        Li$_2$MgH$_{16}$ Cm (300 GPa)            &  {188 } &      {30 }          &-\cite{Sun2019LimgH}\\
        Li$_2$MgH$_{16}$ C2/m (300 GPa)          &  {187 } &      {30 }          &-\cite{Sun2019LimgH}\\
        Li$_2$MgH$_{16}$ Cmc2 (300 GPa)          &  {173 } &      {30 }          &-\cite{Sun2019LimgH}\\
        Li$_2$MgH$_{16}$ {I4 (300 GPa)}            &  {200 } &      {30 }          &-\cite{Sun2019LimgH}\\
        Li$_2$MgH$_{16}$ {R32 (300 GPa)}           &  {259 } &      {30 }          &-\cite{Sun2019LimgH}\\
        LiMgH$_{10}$ R$\bar{3}$m (300 GPa)            &   {267 } &     {30 }          &-\cite{Sun2019LimgH}\\
        LiMgH$_{14}$ Cc (300 GPa)              &  {147 } &      {32 }          &-\cite{Sun2019LimgH}\\
        
        \hline
        
     \end{tabular}
    \caption{\label{Table:Tcpredictionanalysis}Several high temperature superconductors and their predicted $T_c$s through various methods.
    $T_c$ pred is this model's prediction, and $T_c$ chem is the chemical type model prediction.\cite{chemM} The final column is the $T_c$ by previous computations or experiments if available. For calculated $T_c$s the average value of the presented $\mu^*$ domain is used. ${}^*$ The calculation is reproduced for this work using the approach of reference \cite{Novakovic2022} and an adjusted 0.015 Ry Gaussian smearing width.}
    \end{table}
    
    This model provides successful predictions in several studied high temperature superconductors as seen in Table \ref{Table:Tcpredictionanalysis}.
    In general, this study's model predictions are in much better agreement than the composition-based model\cite{chemM} predictions (Table \ref{Table:Tcpredictionanalysis}) which are unable to predict high $T_c$ structures within the pressure structure landscape.
    In particular, we were interested in how well this model could predict the superconducting properties of high pressure hydride materials, so structures for materials such as C-S-H, LaH$_{x}$, and YH$_{x}$ were taken from recently published studies.\cite{Cui2020,Novakovic2022,Shipley2020,Shipley2021,Wang2022,Drozdov2019} 
    Prediction for several of the hydrides are within good agreement of experiment such as a 237\,K $T_c$ prediction for P6$_3$/mmc YH$_9$ at 255 GPa that has a measured $T_c$ of 237 K. \cite{Kong2021} The predicted $T_c$ for LaH$_{10}$ Fm$\bar{3}$m at 150 GPa pressure is also close to the measured value of 249 K.\cite{Drozdov2019} A majority of the computed phases such as LaH$_{10}$, YH$_{10}$, and CSH$_7$ at pressures above 250 GPa are in strong agreement to the calculated $T_c$s through electron-phonon density functional theory (DFT) calculations.\cite{Shipley2020} 
    Recently we published an investigation into the dependence of the chosen density functional on the calculated $T_c$ for the predicted CSH$_7$ superconductor at 270 GPa.\cite{Novakovic2022} It was found there that the vdW corrections to the density functional boosted the estimated $T_c$ of a predicted CSH$_7$ polymorph from 80 K to 174 K, which closely follows the model calculated value of 177 K. 
    These are in comparison to the chemical composition predicted values which did not predict above 37 K and cannot distinguish different pressures and phases by the nature of their description.
    The LiMgH superconductors are not as near to the predicted values though these are the considered an extreme boundary of predicting with electron-phonon calculations as theory is expected to be less reliable for calculation for high temperatures and strong electron-phonon coupling. 
    They are likely more difficult to evaluate through the same system given their $T_c$s are a large outlier in comparison to those of superconductors used in the training data.

\subsection{Morphisms and Discovery}
    
With large systems electron-phonon calculations may take several weeks for accurate $T_c$s through Migdal-Eliashberg theory, which can be estimated quickly by the model.
Given the depth and quickness of the model, it can facilitate the discovery of candidate high temperature superconductors with diverse and complex structures.
        
Sets of theoretical superconductors generated through morphisms of base structures through varying the crystal lattice and swapping atomic species efficiently allows quickly probing an extensive compositional space of diverse superconductors.

Using a system with a low number of atoms as a base is preferable for generating simple structures which are more realistic for measurement and calculation via conventional density functional theory (DFT) approaches, while large systems allows exploration of more complicated landscapes of theoretical high $T_c$ materials where conventional DFT approaches could be cost prohibitive. 
Im$\bar{3}$m H$_3$S at 200 GPa is used as the starting point for the smaller systems because of its simple composition and predicted $\sim$200 K $T_c$ \cite{anharmonich3} near the experimental value of 203\,K.\cite{h3s}
Whereas Fm$\bar{3}$m MgH$_{13}$ supplies a rich space of variants of larger systems to explore owing to its close to 300\,K predicted $T_c$ \cite{Shipley2021} as well as its hydrogen-dense composition alike closely studied high $T_c$ superconductors such as LaH$_{10}$.
  
\begin{figure}[h!]
            \centering
            \includegraphics[width=0.5\textwidth]{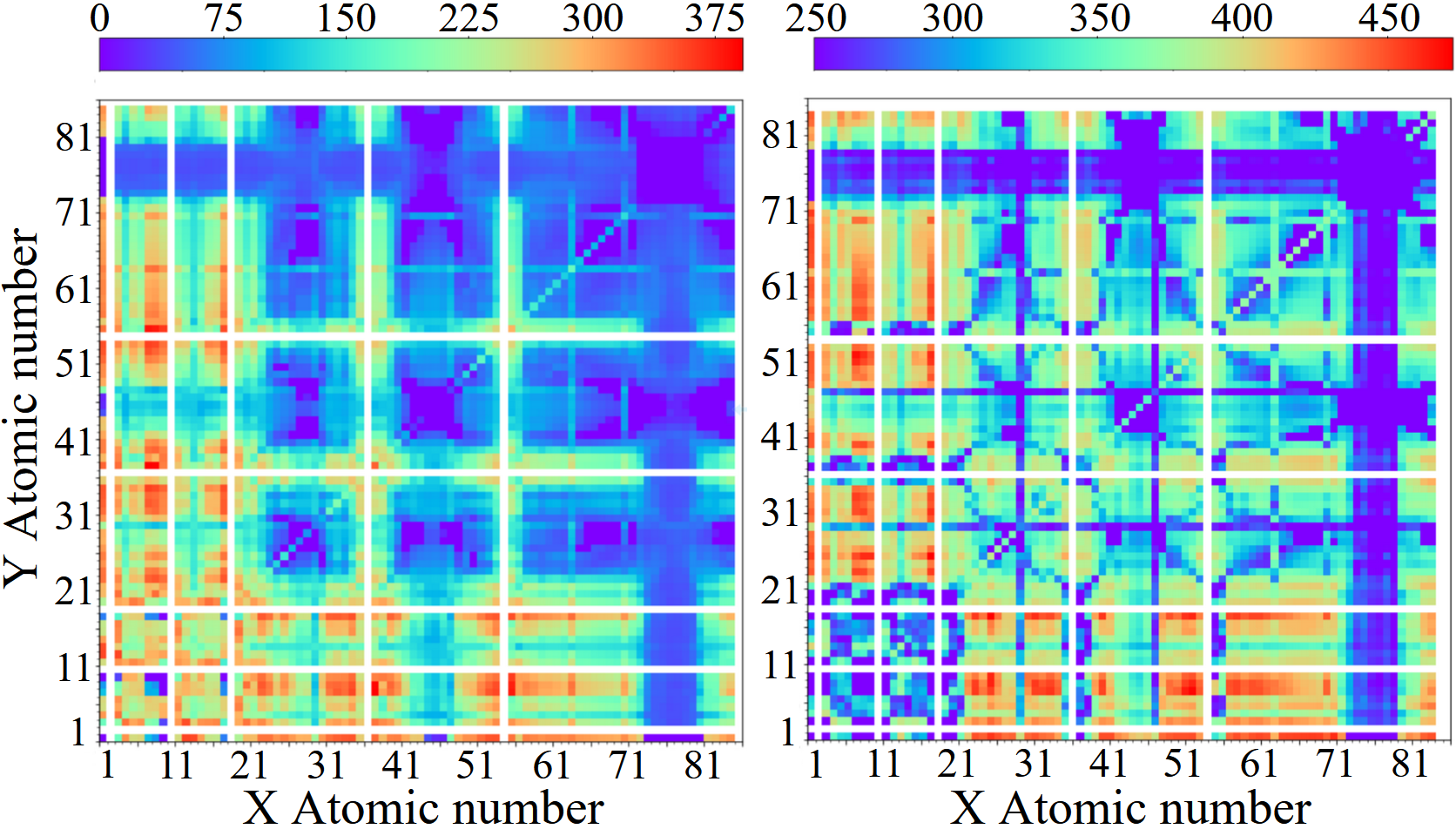}
              \caption{{(left) $T_c$ of Fm$\bar{3}$m MgH$_{13}$ morphisms with X representing elements swapped with the magnesium atom and Y representing elements swapped with the center hydrogen atom. 
              (right) $T_c$ of Fd$\bar{3}$m Li$_2$MgH$_{16}$ morphisms. Here the Li atoms are set to H and the Mg atoms are varied. 
              Atoms with Z=1 to 86 are used, and  combinations with inert elements (He, Ne, etc.) are expressed by white lines as $T_c$ is not predicted. A $T_c$ vs lattice scale $a$ dome is calculated and the peak $T_c$ is presented here.
              The initial parameters for the primitive rhombohedral unit cells of Fm$\bar{3}$m MgH$_{13}$ and Fd$\bar{3}$m Li$_2$MgH$_{16}$ are $a=3.469(5)$\,\AA{} and $a=4.750(7)$\,\AA{} respectively.\cite{Shipley2021,Sun2019LimgH}
              }}

            \label{fig:xyH36}
        \end{figure}

Figure \ref{fig:xyH36} shows $T_c$ map for Fm$\bar{3}$m MgH$_{13}$ (200 GPa) with the Mg and H  \cite{Shipley2021}$^,$\footnote{Structure files are available by \url{https://github.com/LazarNov/superconductor-predict}. Atoms labelled as X/Y specify the morphed sites.} each morphed to atoms X and Y, with atomic numbers between 1 to 86. 
For every generated structure, the lattice parameters were scaled uniformly to determine the respective predicted peak $T_c$. 
From this process Fm$\bar{3}$m LiMgH$_{12}$ and AlH$_{13}$ were among the highest suggested superconductors through this discovery method with maximum predicted $T_c$s of 356 K and 353 K from Table \ref{Table:predictedsuperconductors}. 
LiMgH$_{12}$ falls among a family of theoretically studied Li$_x$MgH$_y$ superconductors, many reported to have very high $T_c$s. Most notably from that study is Fd$\bar{3}$m Li$_2$MgH$_{16}$ which is predicted to have a $T_c$ up to 473 K.\cite{Sun2019LimgH}

        For the sulphur-exchanged variations of H$_3$S, lattices were kept at the original lattice parameters to maintain an idea of the general pressure range. The sulfur-exchanged variants of H$_3$S place Im$\bar{3}$m PH$_3$, BH$_3$, SeH$_3$ in the pool of potential high $T_c$ superconductors with $T_c$s of 191 K, 140 K, and 113 K. These results compare quite well to studied superconductors with Pbcn BH$_3$ predicted to have a $T_c$ of 125 K at 360 GPa\cite{BH3} and a 200 GPa undetermined phase of PH$_3$ having a measured $T_c$ above 100 K.\cite{ph} Remarkably, the same Im$\bar{3}$m SeH$_3$ structure has been estimated through DFT-based approaches to have a nearly identical $T_c$ as what was predicted here, ie. 110 K at 200 GPa.\cite{seh3}
        Following that favorable comparison, our Im$\bar{3}$m SeH$_3$ structure was optimized with Quantum Espresso \cite{Giannozzi_2009,Giannozzi_2017} at 200 GPa using the respective vdW-DF2 optimization procedure as detailed in our recent work.\cite{Novakovic2022} This updated the predicted $T_c$ to 111 K from 113 K.

        Following the encouraging results on LiMgH$_{12}$ result, the predicted 250 GPa Fd$\bar{3}$m Li$_2$MgH$_{16}$ material\cite{Sun2019LimgH}, 2 formula units in the cell,  was used as a base for generating morphisms. Base stoichiometry of the material is therefore Li$_4$Mg$_2$H$_{32}$.
        Given the high number of atoms, the space for discovery was limited to studying interchanging the Li and Mg atoms. The entire space of 6 atom variations is on the order of $86^6 \approx 10^{12}$ variations without symmetry. This number is completely intractable for first principles evaluations, and it still requires preliminary probing and use of further constraints with the model presented here. 
        A shallow random search explored thousands of morphisms in this space, with magnesium atoms and pairs of lithium atoms replaced with random species of atomic number between 1 and 86. This revealed BaH$_{32}$N$_5$ and LaH$_{32}$N$_5$, as well as BaH$_{32}$N$_4$O and LaH$_{32}$N$_4$O  variants, well above others in the random probing as potential high temperature superconductors with $T_c$ $>$ 350 K as shown in Table \ref{Table:predictedsuperconductors}. The number of atoms are fixed using this process and it is possible that by modification, likely the hydrogen density, a lower atom content structure possessing a very high $T_c$ may be generated through these.
        
Hydrogen content replacing lithium atoms showed a higher frequency of high $T_c$ superconductors in the previous example, and it encouraged replacing the Li$_4$ of the Li$_4$Mg$_2$H$_{32}$ to H$_4$ and investigating morphisms of the form XYH$_{36}$, ie. only interchanging the magnesium atoms.
Figure \ref{fig:xyH36} shows the map of $T_c$ for Fd$\bar{3}$m XYH$_{32}$ with all atomic pairs between atomic number 1 and 86.
This morphism yields several potential room temperature superconductors with $T_c$ well above above 400 K such as LiLaH$_{36}$, ZrH$_{36}$Cl, and  TeH$_{36}$N as noted in Table \ref{Table:predictedsuperconductors}.
The composition model does not capture any of these potential superconductors.
Interestingly, the $T_c$s are predicted to remain high above 200 K even for the very wide scale of the lattice size, implying that lower pressure structures potentially conserve high $T_c$ superconductivity.

\begin{table}[h!]
        
            \begin{tabular}{|l|l|l|}
        \hline
        
        {{Superconductor}} & 
        {{$T_c$ pred (K)}}  & {{$T_{c}$ chem (K) }} \\
        \hline
        {LiMgH$_{12}$ Fm$\bar{3}$m}        & {356 - 213}       & {34}\\
        {AlH$_{13}$  Fm$\bar{3}$m}         & {353 - 249}      & {47}\\
        
        {MnH$_{13}$ Fm$\bar{3}$m}          & {327 - 231}      & {25}\\
        \hline
         {BaH$_{32}$N$_5$ Fd$\bar{3}$m}     & {385 - 302}     & {42}\\
        {LaH$_{32}$N$_5$ Fd$\bar{3}$m}     & {434 - 318}     & {41}\\
        {BaH$_{32}$N$_{4}$O Fd$\bar{3}$m}  & {351 - 234}     & {39}\\
        {LaH$_{32}$N$_4$O Fd$\bar{3}$m}    & {384 - 254}    & {38}\\

        \hline
        {TlBH$_{36}$ Fd$\bar{3}$m}         & {403 - 282}   & {26}\\
        {TlH$_{37}$ Fd$\bar{3}$m}          & {432 - 291}  & {38}\\
        
        LuH$_{18}$ Fd$\bar{3}$m &  362 - 243   & 31 \\
        
        LiLaH$_{36}$ Fd$\bar{3}$m &  430 - 316  & 30 \\
        
        ZrH$_{36}$Cl Fd$\bar{3}$m &  457 - 332   & 41 \\
        
        TeH$_{36}$N Fd$\bar{3}$m &  471 - 355   & 46 \\
        \hline
        
        PH$_3$ Im$\bar{3}$m  & 191  & 64 \\
        
        BH$_3$ Im$\bar{3}$m & 140  & 59 \\
        
        SeH$_3$ Im$\bar{3}$m & 113  & 28 \\

        \hline 
        \end{tabular}

        \caption{\label{Table:predictedsuperconductors} Estimated model $T_c$s of Fm$\bar{3}$m XYH$_{12}$ (MgH$_{13}$),  Fd$\bar{3}$m XYH$_{36}$ (Li$_{2}$MgH$_{16}$), and Im$\bar{3}$m XH$_3$ (H$_3$S) morphisms. For XH$_{3}$ structures the lattice parameters kept the 200\,GPa H$_{3}$S primitive rhombohedral lattice values of $a=2.585$ \AA\ and $\alpha = 109.47\degree$. For other superconductors the lattice parameters are uniformly scaled to 6.0 \AA\ and the highest and lowest $T_c$s are expressed. Composition model predictions\cite{chemM} are expressed by column 3, which do not change with structure.} 

\end{table} 
        
    \subsection{Physical Aspects}
    
\begin{figure}[h!]
            \includegraphics[scale=0.1900]{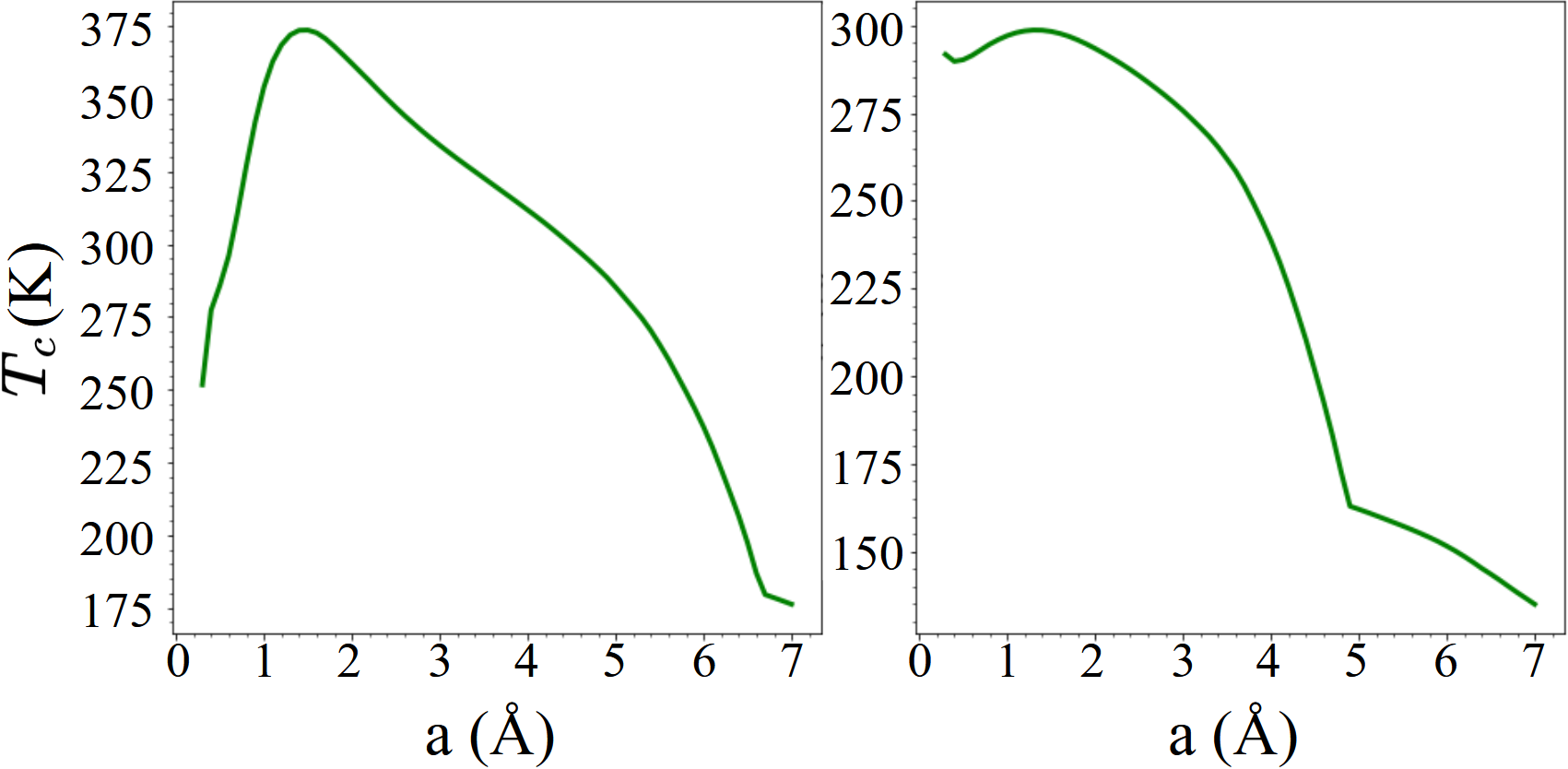}
            \caption{$T_c$ vs. lattice scaling $a$ for the primitive rhombohedral unit cell of  Fd$\bar{3}$m (left) and P$\bar{3}$m1 (right) Li$_2$MgH$_{16}$. Lattice parameters are scaled uniformly with $c/a$ ratios and angles fixed. Initial parameters are $a=4.750(7)$\,\AA{} for the primitive rhombohedral cell of Fd$\bar{3}$m Li$_2$MgH$_{16}$ and $a=2.795(9)$\,\AA{} and $c=5.313(3)$\,\AA{} for P$\bar{3}$m1 Li$_2$MgH$_{16}$. The initial parameters of Fd$\bar{3}$m and P$\bar{3}$m1 Li$_2$MgH$_{16}$ were predicted at 250 and 300\,GPa respectively.\cite{Sun2019LimgH}}

             \label{fig:Tcvsa}
    \end{figure}

    Within the approximation that a material is isomorphic as a function of pressure, varying the lattice parameters 
    over a large range effectively probes the pressure dependent "superconducting dome" of a candidate superconductor. Remarkably, uniformly scaling the lattice of the structures in the data set reveals a superconducting dome similar to $T_c$ versus pressure effects observed experimentally.\cite{Drozdov2019,PhysRevB.105.094507} Polymorphisms had distinct $T_c$ responses to uniform scaling of lattice vectors. This is demonstrated by figure \ref{fig:Tcvsa} where Fd$\bar{3}$m and P$\bar{3}$m1 Li$_2$MgH$_{16}$ show different volume effects on $T_c$. 
    
        \begin{figure}[h!]
            \centering
            \includegraphics[width=1.00\columnwidth]{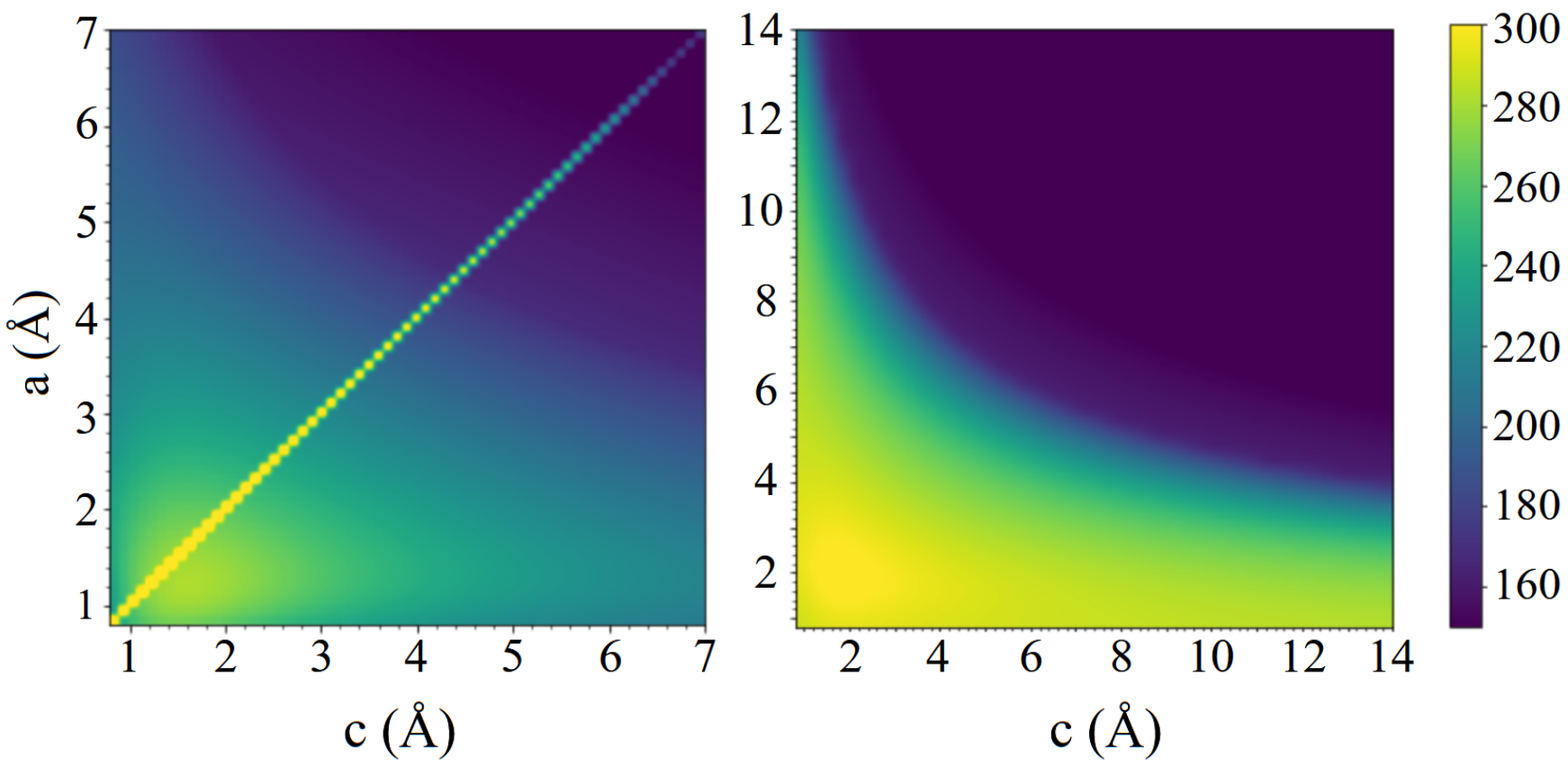}
            \includegraphics[width=1.00\columnwidth]{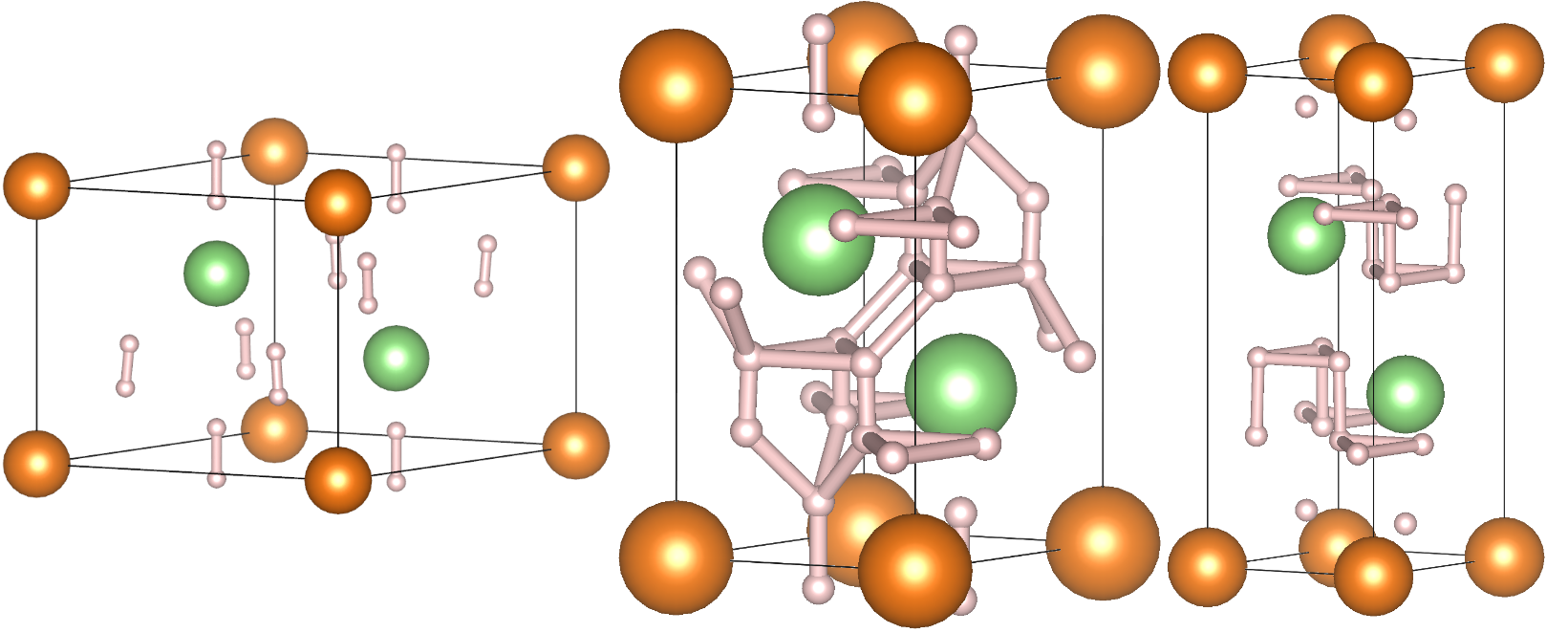}
            \caption{
            The predicted $T_c$ for a space of primitive Fd$\bar{3}$m (upper left) and P$\bar{3}$m1 (upper right) Li$_2$MgH$_{16}$ lattices. The space spans the lattices with $a$ and $c$ independently varying between 0.8\,--\,7.0\,\AA{} for the Fd$\bar{3}$m phase and 0.8\,--\,14.0\,\AA{} for the P$\bar{3}$m1 phase. Lattice angles are unchanged and $b$ is constrained by $a=b$ in both cases. Initial parameters are $a=4.750(7)$\,\AA{} for the primitive rhombohedral Fd$\bar{3}$m cell and $a=2.795(9)$\,\AA{} and $c=5.313(3)$\,\AA{} for P$\bar{3}$m1 (shown bottom center). The $c=8$\,\AA{} (bottom right) and $a=6$\,\AA{} (bottom left) elongations of the initial (bottom center) P$\bar{3}$m1 structure. H--H interaction distances are cut off at 1.4\,\AA.}
            
            \label{fig:acTc}
        \end{figure}

    Varying the lattice parameters $a=b$ and $c$  vs. $T_c$ for the Fd$\bar{3}$m (in it's primitive representation) and P$\bar{3}$m1 structures of Li$_2$MgH$_{12}$ also exhibits different behavior for the different structures shown in figure~\ref{fig:acTc}. The Fd$\bar{3}$m structure peaks in $T_c$ with equal lattice lengths (the ideal rhombohedral cell), and $T_c$ drops significantly with asymmetry.
    The P$\bar{3}$m1 structure on the other hand shows a smooth behavior of the lattice effects on the predicted $T_c$, with a preference for smaller volumes (ie. higher pressures). The P$\bar{3}$m1 structure shows a rapid drop off in $T_c$ for an expanded $a=b$-axis, whereas it is more tolerant for deformations of the $c$-axis. The bottom center structure shown in figure~\ref{fig:acTc} shows the P$\bar{3}$m1 structure with its initial lattice settings of $a=2.795(9)$\,\AA{} and $c=5.313(3)$\,\AA, and from this one can see a somewhat 3-dimensional bonding network if the H--H interaction distances are cut off at 1.4\,\AA{} --- a reasonable distance cutoff for visualizing the clathrate-like H cages in structures such as YH$_6$ and YH$_9$.\cite{Kong2021,Wang2022} When the $c$-axis is stretched (keeping $a=b$ fixed) as shown in the bottom right of figure~\ref{fig:acTc} (with $c=8$\,\AA), the H--H interaction networks in the $ab$ planes are not broken. However, keeping $c$ fixed and stretching $a=b$ as shown with $a=6$\,\AA{} in the bottom left of figure~\ref{fig:acTc} disrupts this 2-dimensional connectivity leaving only non-connected, molecular-like H--H units, strongly implying that the 2-dimensional H--H structural motif is what is driving the $T_c$ predicted by our model.\\

\section{Conclusions}

Composition models have presented a successful concept for possibilities in machine learning approaches to superconductor research. 
However, implementing such models for the analysis and discovery of superconductors requires completely encapsulating their identity. 
Including structural descriptors as is done here enables studying characteristic behavior such as superconducting domes, symmetry, and atomic periodicity which is proving to be critical for the discovery and understanding of superconducting hydride materials at high pressures.
The periodic electron and mass densities descriptors were represented with a complex-valued CNN as their smooth densities are naturally complex valued and are analogous to 3d image data for which CNNs are commonly employed.
Though chemical composition derived properties are built within this density representation, non-structural parameters are still used because of relevant information such as electronegativity and number of valence electrons, which are not expressed formally by densities of mass and charge.

The structural representation independently suggests studied theoretical superconductors such as aluminum and iodine hydrides as well as LiMgH$_{12}$, results not captured by composition models.
The model presented here also provides unique inverses of the atomic structure by the Fourier representation, thus enabling purely $T_c$ based explorations of structural landscapes as highlighted here with the case study on Li$_2$MgH$_{16}$.
Utilizing a structural based machine learned modelling of $T_c$ along with crystal structure prediction can be used in the future to create a screening procedure that will increase the discovery efficiency of experimentally feasible materials.
Especially, as future versions could benefit from more accurate representations of the electron and mass densities derived from CSP.
Theoretical and experimental validation of the structures predicted by this model will only increase data availability which can be used to further refine such machine learning approaches.

This work's data and code will be made available at 
\url{https://github.com/LazarNov/superconductor-predict}.

\section{Acknowledgements}

This work supported by the U.S. Department of Energy, Office of Basic Energy Sciences under Award Number DE-SC0020303.
Computational resources were provided by the UNLV National Supercomputing Institute.

\FloatBarrier
\bibliography{ref.bib}

\begin{thebibliography}{51}%
\makeatletter
\providecommand \@ifxundefined [1]{%
 \@ifx{#1\undefined}
}%
\providecommand \@ifnum [1]{%
 \ifnum #1\expandafter \@firstoftwo
 \else \expandafter \@secondoftwo
 \fi
}%
\providecommand \@ifx [1]{%
 \ifx #1\expandafter \@firstoftwo
 \else \expandafter \@secondoftwo
 \fi
}%
\providecommand \natexlab [1]{#1}%
\providecommand \enquote  [1]{``#1''}%
\providecommand \bibnamefont  [1]{#1}%
\providecommand \bibfnamefont [1]{#1}%
\providecommand \citenamefont [1]{#1}%
\providecommand \href@noop [0]{\@secondoftwo}%
\providecommand \href [0]{\begingroup \@sanitize@url \@href}%
\providecommand \@href[1]{\@@startlink{#1}\@@href}%
\providecommand \@@href[1]{\endgroup#1\@@endlink}%
\providecommand \@sanitize@url [0]{\catcode `\\12\catcode `\$12\catcode
  `\&12\catcode `\#12\catcode `\^12\catcode `\_12\catcode `\%12\relax}%
\providecommand \@@startlink[1]{}%
\providecommand \@@endlink[0]{}%
\providecommand \url  [0]{\begingroup\@sanitize@url \@url }%
\providecommand \@url [1]{\endgroup\@href {#1}{\urlprefix }}%
\providecommand \urlprefix  [0]{URL }%
\providecommand \Eprint [0]{\href }%
\providecommand \doibase [0]{http://dx.doi.org/}%
\providecommand \selectlanguage [0]{\@gobble}%
\providecommand \bibinfo  [0]{\@secondoftwo}%
\providecommand \bibfield  [0]{\@secondoftwo}%
\providecommand \translation [1]{[#1]}%
\providecommand \BibitemOpen [0]{}%
\providecommand \bibitemStop [0]{}%
\providecommand \bibitemNoStop [0]{.\EOS\space}%
\providecommand \EOS [0]{\spacefactor3000\relax}%
\providecommand \BibitemShut  [1]{\csname bibitem#1\endcsname}%
\let\auto@bib@innerbib\@empty
\bibitem [{\citenamefont {Shipley}\ \emph {et~al.}(2021)\citenamefont
  {Shipley}, \citenamefont {Hutcheon}, \citenamefont {Needs},\ and\
  \citenamefont {Pickard}}]{Shipley2021}%
  \BibitemOpen
  \bibfield  {author} {\bibinfo {author} {\bibfnamefont {A.~M.}\ \bibnamefont
  {Shipley}}, \bibinfo {author} {\bibfnamefont {M.~J.}\ \bibnamefont
  {Hutcheon}}, \bibinfo {author} {\bibfnamefont {R.~J.}\ \bibnamefont {Needs}},
  \ and\ \bibinfo {author} {\bibfnamefont {C.~J.}\ \bibnamefont {Pickard}},\
  }\href {\doibase 10.1103/physrevb.104.054501} {\bibfield  {journal} {\bibinfo
   {journal} {Physical Review B}\ }\textbf {\bibinfo {volume} {104}} (\bibinfo
  {year} {2021}),\ 10.1103/physrevb.104.054501}\BibitemShut {NoStop}%
\bibitem [{\citenamefont {Wang}\ \emph
  {et~al.}(2022{\natexlab{a}})\citenamefont {Wang}, \citenamefont {Bi},
  \citenamefont {Hilleke}, \citenamefont {Lamichhane}, \citenamefont {Hemley},\
  and\ \citenamefont {Zurek}}]{dCarbon}%
  \BibitemOpen
  \bibfield  {author} {\bibinfo {author} {\bibfnamefont {X.}~\bibnamefont
  {Wang}}, \bibinfo {author} {\bibfnamefont {T.}~\bibnamefont {Bi}}, \bibinfo
  {author} {\bibfnamefont {K.}~\bibnamefont {Hilleke}}, \bibinfo {author}
  {\bibfnamefont {A.}~\bibnamefont {Lamichhane}}, \bibinfo {author}
  {\bibfnamefont {R.}~\bibnamefont {Hemley}}, \ and\ \bibinfo {author}
  {\bibfnamefont {E.}~\bibnamefont {Zurek}},\ }\href {\doibase
  10.1038/s41524-022-00769-9} {\bibfield  {journal} {\bibinfo  {journal} {npj
  Computational Materials}\ }\textbf {\bibinfo {volume} {8}} (\bibinfo {year}
  {2022}{\natexlab{a}}),\ 10.1038/s41524-022-00769-9}\BibitemShut {NoStop}%
\bibitem [{\citenamefont {Lonie}\ and\ \citenamefont
  {Zurek}(2011)}]{LONIE2011372}%
  \BibitemOpen
  \bibfield  {author} {\bibinfo {author} {\bibfnamefont {D.~C.}\ \bibnamefont
  {Lonie}}\ and\ \bibinfo {author} {\bibfnamefont {E.}~\bibnamefont {Zurek}},\
  }\href {\doibase https://doi.org/10.1016/j.cpc.2010.07.048} {\bibfield
  {journal} {\bibinfo  {journal} {Computer Physics Communications}\ }\textbf
  {\bibinfo {volume} {182}},\ \bibinfo {pages} {372} (\bibinfo {year}
  {2011})}\BibitemShut {NoStop}%
\bibitem [{\citenamefont {Avery}\ \emph {et~al.}(2019)\citenamefont {Avery},
  \citenamefont {Toher}, \citenamefont {Curtarolo},\ and\ \citenamefont
  {Zurek}}]{AVERY2019274}%
  \BibitemOpen
  \bibfield  {author} {\bibinfo {author} {\bibfnamefont {P.}~\bibnamefont
  {Avery}}, \bibinfo {author} {\bibfnamefont {C.}~\bibnamefont {Toher}},
  \bibinfo {author} {\bibfnamefont {S.}~\bibnamefont {Curtarolo}}, \ and\
  \bibinfo {author} {\bibfnamefont {E.}~\bibnamefont {Zurek}},\ }\href
  {\doibase https://doi.org/10.1016/j.cpc.2018.11.016} {\bibfield  {journal}
  {\bibinfo  {journal} {Computer Physics Communications}\ }\textbf {\bibinfo
  {volume} {237}},\ \bibinfo {pages} {274} (\bibinfo {year}
  {2019})}\BibitemShut {NoStop}%
\bibitem [{\citenamefont {Lyakhov}\ \emph {et~al.}(2013)\citenamefont
  {Lyakhov}, \citenamefont {Oganov}, \citenamefont {Stokes},\ and\
  \citenamefont {Zhu}}]{LYAKHOV20131172}%
  \BibitemOpen
  \bibfield  {author} {\bibinfo {author} {\bibfnamefont {A.~O.}\ \bibnamefont
  {Lyakhov}}, \bibinfo {author} {\bibfnamefont {A.~R.}\ \bibnamefont {Oganov}},
  \bibinfo {author} {\bibfnamefont {H.~T.}\ \bibnamefont {Stokes}}, \ and\
  \bibinfo {author} {\bibfnamefont {Q.}~\bibnamefont {Zhu}},\ }\href {\doibase
  https://doi.org/10.1016/j.cpc.2012.12.009} {\bibfield  {journal} {\bibinfo
  {journal} {Comp. Phys. Commun.}\ }\textbf {\bibinfo {volume} {184}},\
  \bibinfo {pages} {1172} (\bibinfo {year} {2013})}\BibitemShut {NoStop}%
\bibitem [{\citenamefont {Wang}\ \emph {et~al.}(2010)\citenamefont {Wang},
  \citenamefont {Lv}, \citenamefont {Zhu},\ and\ \citenamefont {Ma}}]{Calypso}%
  \BibitemOpen
  \bibfield  {author} {\bibinfo {author} {\bibfnamefont {Y.}~\bibnamefont
  {Wang}}, \bibinfo {author} {\bibfnamefont {J.}~\bibnamefont {Lv}}, \bibinfo
  {author} {\bibfnamefont {L.}~\bibnamefont {Zhu}}, \ and\ \bibinfo {author}
  {\bibfnamefont {Y.}~\bibnamefont {Ma}},\ }\href {\doibase
  10.1103/PhysRevB.82.094116} {\bibfield  {journal} {\bibinfo  {journal} {Phys.
  Rev. B}\ }\textbf {\bibinfo {volume} {82}},\ \bibinfo {pages} {094116}
  (\bibinfo {year} {2010})}\BibitemShut {NoStop}%
\bibitem [{\citenamefont {Pickard}\ and\ \citenamefont
  {Needs}(2011)}]{Pickard_2011}%
  \BibitemOpen
  \bibfield  {author} {\bibinfo {author} {\bibfnamefont {C.~J.}\ \bibnamefont
  {Pickard}}\ and\ \bibinfo {author} {\bibfnamefont {R.~J.}\ \bibnamefont
  {Needs}},\ }\href {\doibase 10.1088/0953-8984/23/5/053201} {\bibfield
  {journal} {\bibinfo  {journal} {Journal of Physics: Condensed Matter}\
  }\textbf {\bibinfo {volume} {23}},\ \bibinfo {pages} {053201} (\bibinfo
  {year} {2011})}\BibitemShut {NoStop}%
\bibitem [{\citenamefont {Wierzbowska}\ \emph {et~al.}(2005)\citenamefont
  {Wierzbowska}, \citenamefont {de~Gironcoli},\ and\ \citenamefont
  {Giannozzi}}]{Nbqewmgsgp}%
  \BibitemOpen
  \bibfield  {author} {\bibinfo {author} {\bibfnamefont {M.}~\bibnamefont
  {Wierzbowska}}, \bibinfo {author} {\bibfnamefont {S.}~\bibnamefont
  {de~Gironcoli}}, \ and\ \bibinfo {author} {\bibfnamefont {P.}~\bibnamefont
  {Giannozzi}},\ }\href {\doibase 10.48550/ARXIV.COND-MAT/0504077} {\enquote
  {\bibinfo {title} {Origins of low- and high-pressure discontinuities of
  $t_{c}$ in niobium},}\ } (\bibinfo {year} {2005})\BibitemShut {NoStop}%
\bibitem [{\citenamefont {Shipley}\ \emph
  {et~al.}(2020{\natexlab{a}})\citenamefont {Shipley}, \citenamefont
  {Hutcheon}, \citenamefont {Johnson}, \citenamefont {Needs},\ and\
  \citenamefont {Pickard}}]{sh2020}%
  \BibitemOpen
  \bibfield  {author} {\bibinfo {author} {\bibfnamefont {A.~M.}\ \bibnamefont
  {Shipley}}, \bibinfo {author} {\bibfnamefont {M.~J.}\ \bibnamefont
  {Hutcheon}}, \bibinfo {author} {\bibfnamefont {M.~S.}\ \bibnamefont
  {Johnson}}, \bibinfo {author} {\bibfnamefont {R.~J.}\ \bibnamefont {Needs}},
  \ and\ \bibinfo {author} {\bibfnamefont {C.~J.}\ \bibnamefont {Pickard}},\
  }\href {\doibase 10.1103/PhysRevB.101.224511} {\bibfield  {journal} {\bibinfo
   {journal} {Phys. Rev. B}\ }\textbf {\bibinfo {volume} {101}},\ \bibinfo
  {pages} {224511} (\bibinfo {year} {2020}{\natexlab{a}})}\BibitemShut
  {NoStop}%
\bibitem [{\citenamefont {Giustino}\ \emph {et~al.}(2007)\citenamefont
  {Giustino}, \citenamefont {Cohen},\ and\ \citenamefont {Louie}}]{epw}%
  \BibitemOpen
  \bibfield  {author} {\bibinfo {author} {\bibfnamefont {F.}~\bibnamefont
  {Giustino}}, \bibinfo {author} {\bibfnamefont {M.~L.}\ \bibnamefont {Cohen}},
  \ and\ \bibinfo {author} {\bibfnamefont {S.~G.}\ \bibnamefont {Louie}},\
  }\href {\doibase 10.1103/PhysRevB.76.165108} {\bibfield  {journal} {\bibinfo
  {journal} {Phys. Rev. B}\ }\textbf {\bibinfo {volume} {76}},\ \bibinfo
  {pages} {165108} (\bibinfo {year} {2007})}\BibitemShut {NoStop}%
\bibitem [{\citenamefont {Poncé}\ \emph {et~al.}(2016)\citenamefont {Poncé},
  \citenamefont {Margine}, \citenamefont {Verdi},\ and\ \citenamefont
  {Giustino}}]{PONCE2016116}%
  \BibitemOpen
  \bibfield  {author} {\bibinfo {author} {\bibfnamefont {S.}~\bibnamefont
  {Poncé}}, \bibinfo {author} {\bibfnamefont {E.}~\bibnamefont {Margine}},
  \bibinfo {author} {\bibfnamefont {C.}~\bibnamefont {Verdi}}, \ and\ \bibinfo
  {author} {\bibfnamefont {F.}~\bibnamefont {Giustino}},\ }\href {\doibase
  https://doi.org/10.1016/j.cpc.2016.07.028} {\bibfield  {journal} {\bibinfo
  {journal} {Computer Physics Communications}\ }\textbf {\bibinfo {volume}
  {209}},\ \bibinfo {pages} {116} (\bibinfo {year} {2016})}\BibitemShut
  {NoStop}%
\bibitem [{\citenamefont {Drozdov}\ \emph
  {et~al.}(2015{\natexlab{a}})\citenamefont {Drozdov}, \citenamefont {Eremets},
  \citenamefont {Troyan}, \citenamefont {Ksenofontov},\ and\ \citenamefont
  {Shylin}}]{h3s}%
  \BibitemOpen
  \bibfield  {author} {\bibinfo {author} {\bibfnamefont {A.~P.}\ \bibnamefont
  {Drozdov}}, \bibinfo {author} {\bibfnamefont {M.~I.}\ \bibnamefont
  {Eremets}}, \bibinfo {author} {\bibfnamefont {I.~A.}\ \bibnamefont {Troyan}},
  \bibinfo {author} {\bibfnamefont {V.}~\bibnamefont {Ksenofontov}}, \ and\
  \bibinfo {author} {\bibfnamefont {S.~I.}\ \bibnamefont {Shylin}},\ }\href
  {\doibase 10.1038/nature14964} {\bibfield  {journal} {\bibinfo  {journal}
  {Nature}\ }\textbf {\bibinfo {volume} {525}},\ \bibinfo {pages} {73}
  (\bibinfo {year} {2015}{\natexlab{a}})}\BibitemShut {NoStop}%
\bibitem [{\citenamefont {Drozdov}\ \emph {et~al.}(2019)\citenamefont
  {Drozdov}, \citenamefont {Kong}, \citenamefont {Minkov}, \citenamefont
  {Besedin}, \citenamefont {Kuzovnikov}, \citenamefont {Mozaffari},
  \citenamefont {Balicas}, \citenamefont {Balakirev}, \citenamefont {Graf},
  \citenamefont {Prakapenka}, \citenamefont {Greenberg}, \citenamefont
  {Knyazev}, \citenamefont {Tkacz},\ and\ \citenamefont
  {Eremets}}]{Drozdov2019}%
  \BibitemOpen
  \bibfield  {author} {\bibinfo {author} {\bibfnamefont {A.~P.}\ \bibnamefont
  {Drozdov}}, \bibinfo {author} {\bibfnamefont {P.~P.}\ \bibnamefont {Kong}},
  \bibinfo {author} {\bibfnamefont {V.~S.}\ \bibnamefont {Minkov}}, \bibinfo
  {author} {\bibfnamefont {S.~P.}\ \bibnamefont {Besedin}}, \bibinfo {author}
  {\bibfnamefont {M.~A.}\ \bibnamefont {Kuzovnikov}}, \bibinfo {author}
  {\bibfnamefont {S.}~\bibnamefont {Mozaffari}}, \bibinfo {author}
  {\bibfnamefont {L.}~\bibnamefont {Balicas}}, \bibinfo {author} {\bibfnamefont
  {F.~F.}\ \bibnamefont {Balakirev}}, \bibinfo {author} {\bibfnamefont {D.~E.}\
  \bibnamefont {Graf}}, \bibinfo {author} {\bibfnamefont {V.~B.}\ \bibnamefont
  {Prakapenka}}, \bibinfo {author} {\bibfnamefont {E.}~\bibnamefont
  {Greenberg}}, \bibinfo {author} {\bibfnamefont {D.~A.}\ \bibnamefont
  {Knyazev}}, \bibinfo {author} {\bibfnamefont {M.}~\bibnamefont {Tkacz}}, \
  and\ \bibinfo {author} {\bibfnamefont {M.~I.}\ \bibnamefont {Eremets}},\
  }\href {\doibase 10.1038/s41586-019-1201-8} {\bibfield  {journal} {\bibinfo
  {journal} {Nature}\ }\textbf {\bibinfo {volume} {569}},\ \bibinfo {pages}
  {528} (\bibinfo {year} {2019})}\BibitemShut {NoStop}%
\bibitem [{\citenamefont {Kong}\ \emph {et~al.}(2021)\citenamefont {Kong},
  \citenamefont {Minkov}, \citenamefont {Kuzovnikov}, \citenamefont {Drozdov},
  \citenamefont {Besedin}, \citenamefont {Mozaffari}, \citenamefont {Balicas},
  \citenamefont {Balakirev}, \citenamefont {Prakapenka}, \citenamefont
  {Chariton}, \citenamefont {Knyazev}, \citenamefont {Greenberg},\ and\
  \citenamefont {Eremets}}]{Kong2021}%
  \BibitemOpen
  \bibfield  {author} {\bibinfo {author} {\bibfnamefont {P.}~\bibnamefont
  {Kong}}, \bibinfo {author} {\bibfnamefont {V.~S.}\ \bibnamefont {Minkov}},
  \bibinfo {author} {\bibfnamefont {M.~A.}\ \bibnamefont {Kuzovnikov}},
  \bibinfo {author} {\bibfnamefont {A.~P.}\ \bibnamefont {Drozdov}}, \bibinfo
  {author} {\bibfnamefont {S.~P.}\ \bibnamefont {Besedin}}, \bibinfo {author}
  {\bibfnamefont {S.}~\bibnamefont {Mozaffari}}, \bibinfo {author}
  {\bibfnamefont {L.}~\bibnamefont {Balicas}}, \bibinfo {author} {\bibfnamefont
  {F.~F.}\ \bibnamefont {Balakirev}}, \bibinfo {author} {\bibfnamefont {V.~B.}\
  \bibnamefont {Prakapenka}}, \bibinfo {author} {\bibfnamefont
  {S.}~\bibnamefont {Chariton}}, \bibinfo {author} {\bibfnamefont {D.~A.}\
  \bibnamefont {Knyazev}}, \bibinfo {author} {\bibfnamefont {E.}~\bibnamefont
  {Greenberg}}, \ and\ \bibinfo {author} {\bibfnamefont {M.~I.}\ \bibnamefont
  {Eremets}},\ }\href {\doibase 10.1038/s41467-021-25372-2} {\bibfield
  {journal} {\bibinfo  {journal} {Nature Communications}\ }\textbf {\bibinfo
  {volume} {12}} (\bibinfo {year} {2021}),\
  10.1038/s41467-021-25372-2}\BibitemShut {NoStop}%
\bibitem [{\citenamefont {Somayazulu}\ \emph {et~al.}(2019)\citenamefont
  {Somayazulu}, \citenamefont {Ahart}, \citenamefont {Mishra}, \citenamefont
  {Geballe}, \citenamefont {Baldini}, \citenamefont {Meng}, \citenamefont
  {Struzhkin},\ and\ \citenamefont {Hemley}}]{PhysRevLett.122.027001}%
  \BibitemOpen
  \bibfield  {author} {\bibinfo {author} {\bibfnamefont {M.}~\bibnamefont
  {Somayazulu}}, \bibinfo {author} {\bibfnamefont {M.}~\bibnamefont {Ahart}},
  \bibinfo {author} {\bibfnamefont {A.~K.}\ \bibnamefont {Mishra}}, \bibinfo
  {author} {\bibfnamefont {Z.~M.}\ \bibnamefont {Geballe}}, \bibinfo {author}
  {\bibfnamefont {M.}~\bibnamefont {Baldini}}, \bibinfo {author} {\bibfnamefont
  {Y.}~\bibnamefont {Meng}}, \bibinfo {author} {\bibfnamefont {V.~V.}\
  \bibnamefont {Struzhkin}}, \ and\ \bibinfo {author} {\bibfnamefont {R.~J.}\
  \bibnamefont {Hemley}},\ }\href {\doibase 10.1103/PhysRevLett.122.027001}
  {\bibfield  {journal} {\bibinfo  {journal} {Phys. Rev. Lett.}\ }\textbf
  {\bibinfo {volume} {122}},\ \bibinfo {pages} {027001} (\bibinfo {year}
  {2019})}\BibitemShut {NoStop}%
\bibitem [{\citenamefont {Ma}\ \emph {et~al.}(2022)\citenamefont {Ma},
  \citenamefont {Wang}, \citenamefont {Xie}, \citenamefont {Yang},
  \citenamefont {Wang}, \citenamefont {Zhou}, \citenamefont {Liu},
  \citenamefont {Yu}, \citenamefont {Zhao}, \citenamefont {Wang}, \citenamefont
  {Liu},\ and\ \citenamefont {Ma}}]{Ma2022}%
  \BibitemOpen
  \bibfield  {author} {\bibinfo {author} {\bibfnamefont {L.}~\bibnamefont
  {Ma}}, \bibinfo {author} {\bibfnamefont {K.}~\bibnamefont {Wang}}, \bibinfo
  {author} {\bibfnamefont {Y.}~\bibnamefont {Xie}}, \bibinfo {author}
  {\bibfnamefont {X.}~\bibnamefont {Yang}}, \bibinfo {author} {\bibfnamefont
  {Y.}~\bibnamefont {Wang}}, \bibinfo {author} {\bibfnamefont {M.}~\bibnamefont
  {Zhou}}, \bibinfo {author} {\bibfnamefont {H.}~\bibnamefont {Liu}}, \bibinfo
  {author} {\bibfnamefont {X.}~\bibnamefont {Yu}}, \bibinfo {author}
  {\bibfnamefont {Y.}~\bibnamefont {Zhao}}, \bibinfo {author} {\bibfnamefont
  {H.}~\bibnamefont {Wang}}, \bibinfo {author} {\bibfnamefont {G.}~\bibnamefont
  {Liu}}, \ and\ \bibinfo {author} {\bibfnamefont {Y.}~\bibnamefont {Ma}},\
  }\href {\doibase 10.1103/physrevlett.128.167001} {\bibfield  {journal}
  {\bibinfo  {journal} {Physical Review Letters}\ }\textbf {\bibinfo {volume}
  {128}} (\bibinfo {year} {2022}),\ 10.1103/physrevlett.128.167001}\BibitemShut
  {NoStop}%
\bibitem [{\citenamefont {Monacelli}\ \emph {et~al.}(2021)\citenamefont
  {Monacelli}, \citenamefont {Bianco}, \citenamefont {Cherubini}, \citenamefont
  {Calandra}, \citenamefont {Errea},\ and\ \citenamefont
  {Mauri}}]{SSCHA-Monacelli_2021}%
  \BibitemOpen
  \bibfield  {author} {\bibinfo {author} {\bibfnamefont {L.}~\bibnamefont
  {Monacelli}}, \bibinfo {author} {\bibfnamefont {R.}~\bibnamefont {Bianco}},
  \bibinfo {author} {\bibfnamefont {M.}~\bibnamefont {Cherubini}}, \bibinfo
  {author} {\bibfnamefont {M.}~\bibnamefont {Calandra}}, \bibinfo {author}
  {\bibfnamefont {I.}~\bibnamefont {Errea}}, \ and\ \bibinfo {author}
  {\bibfnamefont {F.}~\bibnamefont {Mauri}},\ }\href {\doibase
  10.1088/1361-648x/ac066b} {\bibfield  {journal} {\bibinfo  {journal} {Journal
  of Physics: Condensed Matter}\ }\textbf {\bibinfo {volume} {33}},\ \bibinfo
  {pages} {363001} (\bibinfo {year} {2021})}\BibitemShut {NoStop}%
\bibitem [{\citenamefont {Novakovic}\ \emph {et~al.}(2022)\citenamefont
  {Novakovic}, \citenamefont {Sayre}, \citenamefont {Schacher}, \citenamefont
  {Dias}, \citenamefont {Salamat},\ and\ \citenamefont
  {Lawler}}]{Novakovic2022}%
  \BibitemOpen
  \bibfield  {author} {\bibinfo {author} {\bibfnamefont {L.}~\bibnamefont
  {Novakovic}}, \bibinfo {author} {\bibfnamefont {D.}~\bibnamefont {Sayre}},
  \bibinfo {author} {\bibfnamefont {D.}~\bibnamefont {Schacher}}, \bibinfo
  {author} {\bibfnamefont {R.~P.}\ \bibnamefont {Dias}}, \bibinfo {author}
  {\bibfnamefont {A.}~\bibnamefont {Salamat}}, \ and\ \bibinfo {author}
  {\bibfnamefont {K.~V.}\ \bibnamefont {Lawler}},\ }\href {\doibase
  10.1103/physrevb.105.024512} {\bibfield  {journal} {\bibinfo  {journal}
  {Physical Review B}\ }\textbf {\bibinfo {volume} {105}} (\bibinfo {year}
  {2022}),\ 10.1103/physrevb.105.024512}\BibitemShut {NoStop}%
\bibitem [{\citenamefont {Shipley}\ \emph
  {et~al.}(2020{\natexlab{b}})\citenamefont {Shipley}, \citenamefont
  {Hutcheon}, \citenamefont {Johnson}, \citenamefont {Needs},\ and\
  \citenamefont {Pickard}}]{Shipley2020}%
  \BibitemOpen
  \bibfield  {author} {\bibinfo {author} {\bibfnamefont {A.~M.}\ \bibnamefont
  {Shipley}}, \bibinfo {author} {\bibfnamefont {M.~J.}\ \bibnamefont
  {Hutcheon}}, \bibinfo {author} {\bibfnamefont {M.~S.}\ \bibnamefont
  {Johnson}}, \bibinfo {author} {\bibfnamefont {R.~J.}\ \bibnamefont {Needs}},
  \ and\ \bibinfo {author} {\bibfnamefont {C.~J.}\ \bibnamefont {Pickard}},\
  }\href {\doibase 10.1103/physrevb.101.224511} {\bibfield  {journal} {\bibinfo
   {journal} {Physical Review B}\ }\textbf {\bibinfo {volume} {101}} (\bibinfo
  {year} {2020}{\natexlab{b}}),\ 10.1103/physrevb.101.224511}\BibitemShut
  {NoStop}%
\bibitem [{\citenamefont {Snider}\ \emph {et~al.}(2020)\citenamefont {Snider},
  \citenamefont {Dasenbrock-Gammon}, \citenamefont {McBride}, \citenamefont
  {Debessai}, \citenamefont {Vindana}, \citenamefont {Vencatasamy},
  \citenamefont {Lawler}, \citenamefont {Salamat},\ and\ \citenamefont
  {Dias}}]{Snider2020}%
  \BibitemOpen
  \bibfield  {author} {\bibinfo {author} {\bibfnamefont {E.}~\bibnamefont
  {Snider}}, \bibinfo {author} {\bibfnamefont {N.}~\bibnamefont
  {Dasenbrock-Gammon}}, \bibinfo {author} {\bibfnamefont {R.}~\bibnamefont
  {McBride}}, \bibinfo {author} {\bibfnamefont {M.}~\bibnamefont {Debessai}},
  \bibinfo {author} {\bibfnamefont {H.}~\bibnamefont {Vindana}}, \bibinfo
  {author} {\bibfnamefont {K.}~\bibnamefont {Vencatasamy}}, \bibinfo {author}
  {\bibfnamefont {K.~V.}\ \bibnamefont {Lawler}}, \bibinfo {author}
  {\bibfnamefont {A.}~\bibnamefont {Salamat}}, \ and\ \bibinfo {author}
  {\bibfnamefont {R.~P.}\ \bibnamefont {Dias}},\ }\href {\doibase
  10.1038/s41586-020-2801-z} {\bibfield  {journal} {\bibinfo  {journal}
  {Nature}\ }\textbf {\bibinfo {volume} {586}},\ \bibinfo {pages} {373}
  (\bibinfo {year} {2020})}\BibitemShut {NoStop}%
\bibitem [{\citenamefont {Cataldo}\ \emph {et~al.}(2021)\citenamefont
  {Cataldo}, \citenamefont {Heil}, \citenamefont {von~der Linden},\ and\
  \citenamefont {Boeri}}]{DiCataldo2021}%
  \BibitemOpen
  \bibfield  {author} {\bibinfo {author} {\bibfnamefont {S.~D.}\ \bibnamefont
  {Cataldo}}, \bibinfo {author} {\bibfnamefont {C.}~\bibnamefont {Heil}},
  \bibinfo {author} {\bibfnamefont {W.}~\bibnamefont {von~der Linden}}, \ and\
  \bibinfo {author} {\bibfnamefont {L.}~\bibnamefont {Boeri}},\ }\href
  {\doibase 10.1103/physrevb.104.l020511} {\bibfield  {journal} {\bibinfo
  {journal} {Physical Review B}\ }\textbf {\bibinfo {volume} {104}} (\bibinfo
  {year} {2021}),\ 10.1103/physrevb.104.l020511}\BibitemShut {NoStop}%
\bibitem [{\citenamefont {Sun}\ \emph {et~al.}(2019)\citenamefont {Sun},
  \citenamefont {Lv}, \citenamefont {Xie}, \citenamefont {Liu},\ and\
  \citenamefont {Ma}}]{Sun2019LimgH}%
  \BibitemOpen
  \bibfield  {author} {\bibinfo {author} {\bibfnamefont {Y.}~\bibnamefont
  {Sun}}, \bibinfo {author} {\bibfnamefont {J.}~\bibnamefont {Lv}}, \bibinfo
  {author} {\bibfnamefont {Y.}~\bibnamefont {Xie}}, \bibinfo {author}
  {\bibfnamefont {H.}~\bibnamefont {Liu}}, \ and\ \bibinfo {author}
  {\bibfnamefont {Y.}~\bibnamefont {Ma}},\ }\href {\doibase
  10.1103/physrevlett.123.097001} {\bibfield  {journal} {\bibinfo  {journal}
  {Physical Review Letters}\ }\textbf {\bibinfo {volume} {123}} (\bibinfo
  {year} {2019}),\ 10.1103/physrevlett.123.097001}\BibitemShut {NoStop}%
\bibitem [{\citenamefont {Liu}\ \emph {et~al.}(2017{\natexlab{a}})\citenamefont
  {Liu}, \citenamefont {Sun}, \citenamefont {Wang},\ and\ \citenamefont
  {Lu}}]{Liu2017}%
  \BibitemOpen
  \bibfield  {author} {\bibinfo {author} {\bibfnamefont {L.-L.}\ \bibnamefont
  {Liu}}, \bibinfo {author} {\bibfnamefont {H.-J.}\ \bibnamefont {Sun}},
  \bibinfo {author} {\bibfnamefont {C.~Z.}\ \bibnamefont {Wang}}, \ and\
  \bibinfo {author} {\bibfnamefont {W.-C.}\ \bibnamefont {Lu}},\ }\href
  {\doibase 10.1088/1361-648X/aa787d} {\bibfield  {journal} {\bibinfo
  {journal} {Journal of Physics: Condensed Matter}\ }\textbf {\bibinfo {volume}
  {29}},\ \bibinfo {pages} {325401} (\bibinfo {year}
  {2017}{\natexlab{a}})}\BibitemShut {NoStop}%
\bibitem [{\citenamefont {Liu}\ \emph {et~al.}(2017{\natexlab{b}})\citenamefont
  {Liu}, \citenamefont {Naumov}, \citenamefont {Hoffmann}, \citenamefont
  {Ashcroft},\ and\ \citenamefont {Hemley}}]{liu2017potential}%
  \BibitemOpen
  \bibfield  {author} {\bibinfo {author} {\bibfnamefont {H.}~\bibnamefont
  {Liu}}, \bibinfo {author} {\bibfnamefont {I.~I.}\ \bibnamefont {Naumov}},
  \bibinfo {author} {\bibfnamefont {R.}~\bibnamefont {Hoffmann}}, \bibinfo
  {author} {\bibfnamefont {N.}~\bibnamefont {Ashcroft}}, \ and\ \bibinfo
  {author} {\bibfnamefont {R.~J.}\ \bibnamefont {Hemley}},\ }\href {\doibase
  10.1073/pnas.1704505114} {\bibfield  {journal} {\bibinfo  {journal}
  {Proceedings of the National Academy of Sciences}\ }\textbf {\bibinfo
  {volume} {114}},\ \bibinfo {pages} {6990} (\bibinfo {year}
  {2017}{\natexlab{b}})}\BibitemShut {NoStop}%
\bibitem [{sup()}]{supercon.nims.go.jp}%
  \BibitemOpen
  \href {https://supercon.nims.go.jp/} {\enquote {\bibinfo {title} {Supercon
  database: https://supercon.nims.go.jp},}\ }\BibitemShut {NoStop}%
\bibitem [{\citenamefont {Hamidieh}(2018)}]{chemM}%
  \BibitemOpen
  \bibfield  {author} {\bibinfo {author} {\bibfnamefont {K.}~\bibnamefont
  {Hamidieh}},\ }\href {\doibase 10.48550/ARXIV.1803.10260} {\  (\bibinfo
  {year} {2018}),\ 10.48550/ARXIV.1803.10260}\BibitemShut {NoStop}%
\bibitem [{\citenamefont {Stanev}\ \emph {et~al.}(2018)\citenamefont {Stanev},
  \citenamefont {Oses}, \citenamefont {Kusne}, \citenamefont {Rodriguez},
  \citenamefont {Paglione}, \citenamefont {Curtarolo},\ and\ \citenamefont
  {Takeuchi}}]{Stanev2018}%
  \BibitemOpen
  \bibfield  {author} {\bibinfo {author} {\bibfnamefont {V.}~\bibnamefont
  {Stanev}}, \bibinfo {author} {\bibfnamefont {C.}~\bibnamefont {Oses}},
  \bibinfo {author} {\bibfnamefont {A.~G.}\ \bibnamefont {Kusne}}, \bibinfo
  {author} {\bibfnamefont {E.}~\bibnamefont {Rodriguez}}, \bibinfo {author}
  {\bibfnamefont {J.}~\bibnamefont {Paglione}}, \bibinfo {author}
  {\bibfnamefont {S.}~\bibnamefont {Curtarolo}}, \ and\ \bibinfo {author}
  {\bibfnamefont {I.}~\bibnamefont {Takeuchi}},\ }\href {\doibase
  10.1038/s41524-018-0085-8} {\bibfield  {journal} {\bibinfo  {journal} {npj
  Computational Materials}\ }\textbf {\bibinfo {volume} {4}} (\bibinfo {year}
  {2018}),\ 10.1038/s41524-018-0085-8}\BibitemShut {NoStop}%
\bibitem [{\citenamefont {De}\ \emph {et~al.}(2016)\citenamefont {De},
  \citenamefont {Bart{\'{o} }k}, \citenamefont {Cs{\'{a}}nyi},\ and\
  \citenamefont {Ceriotti}}]{soap1}%
  \BibitemOpen
  \bibfield  {author} {\bibinfo {author} {\bibfnamefont {S.}~\bibnamefont
  {De}}, \bibinfo {author} {\bibfnamefont {A.~P.}\ \bibnamefont {Bart{\'{o}
  }k}}, \bibinfo {author} {\bibfnamefont {G.}~\bibnamefont {Cs{\'{a}}nyi}}, \
  and\ \bibinfo {author} {\bibfnamefont {M.}~\bibnamefont {Ceriotti}},\ }\href
  {\doibase 10.1039/c6cp00415f} {\bibfield  {journal} {\bibinfo  {journal}
  {Physical Chemistry Chemical Physics}\ }\textbf {\bibinfo {volume} {18}},\
  \bibinfo {pages} {13754} (\bibinfo {year} {2016})}\BibitemShut {NoStop}%
\bibitem [{\citenamefont {Jäger}\ \emph {et~al.}(2018)\citenamefont {Jäger},
  \citenamefont {Morooka}, \citenamefont {Canova}, \citenamefont {Himanen},\
  and\ \citenamefont {Foster}}]{soap2}%
  \BibitemOpen
  \bibfield  {author} {\bibinfo {author} {\bibfnamefont {M.}~\bibnamefont
  {Jäger}}, \bibinfo {author} {\bibfnamefont {E.}~\bibnamefont {Morooka}},
  \bibinfo {author} {\bibfnamefont {F.}~\bibnamefont {Canova}}, \bibinfo
  {author} {\bibfnamefont {L.}~\bibnamefont {Himanen}}, \ and\ \bibinfo
  {author} {\bibfnamefont {A.}~\bibnamefont {Foster}},\ }\href {\doibase
  10.1038/s41524-018-0096-5} {\bibfield  {journal} {\bibinfo  {journal} {npj
  Computational Materials}\ }\textbf {\bibinfo {volume} {4}} (\bibinfo {year}
  {2018}),\ 10.1038/s41524-018-0096-5}\BibitemShut {NoStop}%
\bibitem [{\citenamefont {Bart{\'{o} }k}\ \emph {et~al.}(2013)\citenamefont
  {Bart{\'{o} }k}, \citenamefont {Kondor},\ and\ \citenamefont
  {Cs{\'{a}}nyi}}]{soap3}%
  \BibitemOpen
  \bibfield  {author} {\bibinfo {author} {\bibfnamefont {A.~P.}\ \bibnamefont
  {Bart{\'{o} }k}}, \bibinfo {author} {\bibfnamefont {R.}~\bibnamefont
  {Kondor}}, \ and\ \bibinfo {author} {\bibfnamefont {G.}~\bibnamefont
  {Cs{\'{a}}nyi}},\ }\href {\doibase 10.1103/physrevb.87.184115} {\bibfield
  {journal} {\bibinfo  {journal} {Physical Review B}\ }\textbf {\bibinfo
  {volume} {87}} (\bibinfo {year} {2013}),\
  10.1103/physrevb.87.184115}\BibitemShut {NoStop}%
\bibitem [{\citenamefont {Musil}\ \emph {et~al.}(2021)\citenamefont {Musil},
  \citenamefont {Grisafi}, \citenamefont {Bartók}, \citenamefont {Ortner},
  \citenamefont {Csányi},\ and\ \citenamefont
  {Ceriotti}}]{doi:10.1021/acs.chemrev.1c00021}%
  \BibitemOpen
  \bibfield  {author} {\bibinfo {author} {\bibfnamefont {F.}~\bibnamefont
  {Musil}}, \bibinfo {author} {\bibfnamefont {A.}~\bibnamefont {Grisafi}},
  \bibinfo {author} {\bibfnamefont {A.~P.}\ \bibnamefont {Bartók}}, \bibinfo
  {author} {\bibfnamefont {C.}~\bibnamefont {Ortner}}, \bibinfo {author}
  {\bibfnamefont {G.}~\bibnamefont {Csányi}}, \ and\ \bibinfo {author}
  {\bibfnamefont {M.}~\bibnamefont {Ceriotti}},\ }\href {\doibase
  10.1021/acs.chemrev.1c00021} {\bibfield  {journal} {\bibinfo  {journal}
  {Chemical Reviews}\ }\textbf {\bibinfo {volume} {121}},\ \bibinfo {pages}
  {9759} (\bibinfo {year} {2021})},\ \bibinfo {note} {pMID: 34310133},\ \Eprint
  {http://arxiv.org/abs/https://doi.org/10.1021/acs.chemrev.1c00021}
  {https://doi.org/10.1021/acs.chemrev.1c00021} \BibitemShut {NoStop}%
\bibitem [{\citenamefont {Townsend}\ \emph {et~al.}(2020)\citenamefont
  {Townsend}, \citenamefont {Micucci}, \citenamefont {Hymel}, \citenamefont
  {Maroulas},\ and\ \citenamefont {Vogiatzis}}]{Townsend2020}%
  \BibitemOpen
  \bibfield  {author} {\bibinfo {author} {\bibfnamefont {J.}~\bibnamefont
  {Townsend}}, \bibinfo {author} {\bibfnamefont {C.~P.}\ \bibnamefont
  {Micucci}}, \bibinfo {author} {\bibfnamefont {J.~H.}\ \bibnamefont {Hymel}},
  \bibinfo {author} {\bibfnamefont {V.}~\bibnamefont {Maroulas}}, \ and\
  \bibinfo {author} {\bibfnamefont {K.~D.}\ \bibnamefont {Vogiatzis}},\ }\href
  {\doibase 10.1038/s41467-020-17035-5} {\bibfield  {journal} {\bibinfo
  {journal} {Nature Communications}\ }\textbf {\bibinfo {volume} {11}}
  (\bibinfo {year} {2020}),\ 10.1038/s41467-020-17035-5}\BibitemShut {NoStop}%
\bibitem [{\citenamefont {Zhang}\ \emph {et~al.}(2022)\citenamefont {Zhang},
  \citenamefont {Zhu}, \citenamefont {Xiang}, \citenamefont {Zhang},
  \citenamefont {Huang}, \citenamefont {Zhong}, \citenamefont {Qiu},
  \citenamefont {Hu},\ and\ \citenamefont {Lin}}]{soap.model}%
  \BibitemOpen
  \bibfield  {author} {\bibinfo {author} {\bibfnamefont {J.}~\bibnamefont
  {Zhang}}, \bibinfo {author} {\bibfnamefont {Z.}~\bibnamefont {Zhu}}, \bibinfo
  {author} {\bibfnamefont {X.-D.}\ \bibnamefont {Xiang}}, \bibinfo {author}
  {\bibfnamefont {K.}~\bibnamefont {Zhang}}, \bibinfo {author} {\bibfnamefont
  {S.}~\bibnamefont {Huang}}, \bibinfo {author} {\bibfnamefont
  {C.}~\bibnamefont {Zhong}}, \bibinfo {author} {\bibfnamefont {H.-J.}\
  \bibnamefont {Qiu}}, \bibinfo {author} {\bibfnamefont {K.}~\bibnamefont
  {Hu}}, \ and\ \bibinfo {author} {\bibfnamefont {X.}~\bibnamefont {Lin}},\
  }\href {\doibase 10.1021/acs.jpcc.2c01904} {\bibfield  {journal} {\bibinfo
  {journal} {The Journal of Physical Chemistry C}\ }\textbf {\bibinfo {volume}
  {126}},\ \bibinfo {pages} {8922} (\bibinfo {year} {2022})},\ \Eprint
  {http://arxiv.org/abs/https://doi.org/10.1021/acs.jpcc.2c01904}
  {https://doi.org/10.1021/acs.jpcc.2c01904} \BibitemShut {NoStop}%
\bibitem [{\citenamefont {Allen}\ and\ \citenamefont
  {Mitrović}(1983)}]{ALLEN19831}%
  \BibitemOpen
  \bibfield  {author} {\bibinfo {author} {\bibfnamefont {P.~B.}\ \bibnamefont
  {Allen}}\ and\ \bibinfo {author} {\bibfnamefont {B.}~\bibnamefont
  {Mitrović}}\ }(\bibinfo  {publisher} {Academic Press},\ \bibinfo {year}
  {1983})\ pp.\ \bibinfo {pages} {1--92}\BibitemShut {NoStop}%
\bibitem [{\citenamefont {Jain}\ \emph {et~al.}(2013)\citenamefont {Jain},
  \citenamefont {Ong}, \citenamefont {Hautier}, \citenamefont {Chen},
  \citenamefont {Richards}, \citenamefont {Dacek}, \citenamefont {Cholia},
  \citenamefont {Gunter}, \citenamefont {Skinner}, \citenamefont {Ceder},\ and\
  \citenamefont {Persson}}]{matp}%
  \BibitemOpen
  \bibfield  {author} {\bibinfo {author} {\bibfnamefont {A.}~\bibnamefont
  {Jain}}, \bibinfo {author} {\bibfnamefont {S.~P.}\ \bibnamefont {Ong}},
  \bibinfo {author} {\bibfnamefont {G.}~\bibnamefont {Hautier}}, \bibinfo
  {author} {\bibfnamefont {W.}~\bibnamefont {Chen}}, \bibinfo {author}
  {\bibfnamefont {W.~D.}\ \bibnamefont {Richards}}, \bibinfo {author}
  {\bibfnamefont {S.}~\bibnamefont {Dacek}}, \bibinfo {author} {\bibfnamefont
  {S.}~\bibnamefont {Cholia}}, \bibinfo {author} {\bibfnamefont
  {D.}~\bibnamefont {Gunter}}, \bibinfo {author} {\bibfnamefont
  {D.}~\bibnamefont {Skinner}}, \bibinfo {author} {\bibfnamefont
  {G.}~\bibnamefont {Ceder}}, \ and\ \bibinfo {author} {\bibfnamefont {K.~a.}\
  \bibnamefont {Persson}},\ }\href {\doibase 10.1063/1.4812323} {\bibfield
  {journal} {\bibinfo  {journal} {APL Materials}\ }\textbf {\bibinfo {volume}
  {1}},\ \bibinfo {pages} {011002} (\bibinfo {year} {2013})}\BibitemShut
  {NoStop}%
\bibitem [{\citenamefont {Ong}\ \emph {et~al.}(2015)\citenamefont {Ong},
  \citenamefont {Cholia}, \citenamefont {Jain}, \citenamefont {Brafman},
  \citenamefont {Gunter}, \citenamefont {Ceder},\ and\ \citenamefont
  {Persson}}]{Ong_2015}%
  \BibitemOpen
  \bibfield  {author} {\bibinfo {author} {\bibfnamefont {S.~P.}\ \bibnamefont
  {Ong}}, \bibinfo {author} {\bibfnamefont {S.}~\bibnamefont {Cholia}},
  \bibinfo {author} {\bibfnamefont {A.}~\bibnamefont {Jain}}, \bibinfo {author}
  {\bibfnamefont {M.}~\bibnamefont {Brafman}}, \bibinfo {author} {\bibfnamefont
  {D.}~\bibnamefont {Gunter}}, \bibinfo {author} {\bibfnamefont
  {G.}~\bibnamefont {Ceder}}, \ and\ \bibinfo {author} {\bibfnamefont {K.~A.}\
  \bibnamefont {Persson}},\ }\href {\doibase 10.1016/j.commatsci.2014.10.037}
  {\bibfield  {journal} {\bibinfo  {journal} {Computational Materials Science}\
  }\textbf {\bibinfo {volume} {97}},\ \bibinfo {pages} {209} (\bibinfo {year}
  {2015})}\BibitemShut {NoStop}%
\bibitem [{\citenamefont {Ong}\ \emph {et~al.}(2013)\citenamefont {Ong},
  \citenamefont {Richards}, \citenamefont {Jain}, \citenamefont {Hautier},
  \citenamefont {Kocher}, \citenamefont {Cholia}, \citenamefont {Gunter},
  \citenamefont {Chevrier}, \citenamefont {Persson},\ and\ \citenamefont
  {Ceder}}]{pymatgen1}%
  \BibitemOpen
  \bibfield  {author} {\bibinfo {author} {\bibfnamefont {S.~P.}\ \bibnamefont
  {Ong}}, \bibinfo {author} {\bibfnamefont {W.~D.}\ \bibnamefont {Richards}},
  \bibinfo {author} {\bibfnamefont {A.}~\bibnamefont {Jain}}, \bibinfo {author}
  {\bibfnamefont {G.}~\bibnamefont {Hautier}}, \bibinfo {author} {\bibfnamefont
  {M.}~\bibnamefont {Kocher}}, \bibinfo {author} {\bibfnamefont
  {S.}~\bibnamefont {Cholia}}, \bibinfo {author} {\bibfnamefont
  {D.}~\bibnamefont {Gunter}}, \bibinfo {author} {\bibfnamefont {V.~L.}\
  \bibnamefont {Chevrier}}, \bibinfo {author} {\bibfnamefont {K.~A.}\
  \bibnamefont {Persson}}, \ and\ \bibinfo {author} {\bibfnamefont
  {G.}~\bibnamefont {Ceder}},\ }\href {\doibase
  https://doi.org/10.1016/j.commatsci.2012.10.028} {\bibfield  {journal}
  {\bibinfo  {journal} {Computational Materials Science}\ }\textbf {\bibinfo
  {volume} {68}},\ \bibinfo {pages} {314} (\bibinfo {year} {2013})}\BibitemShut
  {NoStop}%
\bibitem [{\citenamefont {LeCun}\ \emph {et~al.}(2015)\citenamefont {LeCun},
  \citenamefont {Bengio},\ and\ \citenamefont {Hinton}}]{LeCun2015DeepL}%
  \BibitemOpen
  \bibfield  {author} {\bibinfo {author} {\bibfnamefont {Y.}~\bibnamefont
  {LeCun}}, \bibinfo {author} {\bibfnamefont {Y.}~\bibnamefont {Bengio}}, \
  and\ \bibinfo {author} {\bibfnamefont {G.}~\bibnamefont {Hinton}},\
  }\href@noop {} {\bibfield  {journal} {\bibinfo  {journal} {Nature}\ }\textbf
  {\bibinfo {volume} {521}},\ \bibinfo {pages} {436} (\bibinfo {year}
  {2015})}\BibitemShut {NoStop}%
\bibitem [{\citenamefont {Barrachina}(2022)}]{cvnn}%
  \BibitemOpen
  \bibfield  {author} {\bibinfo {author} {\bibfnamefont {J.~A.}\ \bibnamefont
  {Barrachina}},\ }\href {\doibase 10.5281/zenodo.7303587} {\enquote {\bibinfo
  {title} {Negu93/cvnn: Complex-valued neural networks},}\ } (\bibinfo {year}
  {2022})\BibitemShut {NoStop}%
\bibitem [{\citenamefont {Abadi}\ \emph {et~al.}(2015)\citenamefont {Abadi},
  \citenamefont {Agarwal}, \citenamefont {Barham}, \citenamefont {Brevdo},
  \citenamefont {Chen}, \citenamefont {Citro}, \citenamefont {Corrado},
  \citenamefont {Davis}, \citenamefont {Dean}, \citenamefont {Devin},
  \citenamefont {Ghemawat}, \citenamefont {Goodfellow}, \citenamefont {Harp},
  \citenamefont {Irving}, \citenamefont {Isard}, \citenamefont {Jia},
  \citenamefont {Jozefowicz}, \citenamefont {Kaiser}, \citenamefont {Kudlur},
  \citenamefont {Levenberg}, \citenamefont {Man\'{e}}, \citenamefont {Monga},
  \citenamefont {Moore}, \citenamefont {Murray}, \citenamefont {Olah},
  \citenamefont {Schuster}, \citenamefont {Shlens}, \citenamefont {Steiner},
  \citenamefont {Sutskever}, \citenamefont {Talwar}, \citenamefont {Tucker},
  \citenamefont {Vanhoucke}, \citenamefont {Vasudevan}, \citenamefont
  {Vi\'{e}gas}, \citenamefont {Vinyals}, \citenamefont {Warden}, \citenamefont
  {Wattenberg}, \citenamefont {Wicke}, \citenamefont {Yu},\ and\ \citenamefont
  {Zheng}}]{tensorflow2015-whitepaper}%
  \BibitemOpen
  \bibfield  {author} {\bibinfo {author} {\bibfnamefont {M.}~\bibnamefont
  {Abadi}}, \bibinfo {author} {\bibfnamefont {A.}~\bibnamefont {Agarwal}},
  \bibinfo {author} {\bibfnamefont {P.}~\bibnamefont {Barham}}, \bibinfo
  {author} {\bibfnamefont {E.}~\bibnamefont {Brevdo}}, \bibinfo {author}
  {\bibfnamefont {Z.}~\bibnamefont {Chen}}, \bibinfo {author} {\bibfnamefont
  {C.}~\bibnamefont {Citro}}, \bibinfo {author} {\bibfnamefont {G.~S.}\
  \bibnamefont {Corrado}}, \bibinfo {author} {\bibfnamefont {A.}~\bibnamefont
  {Davis}}, \bibinfo {author} {\bibfnamefont {J.}~\bibnamefont {Dean}},
  \bibinfo {author} {\bibfnamefont {M.}~\bibnamefont {Devin}}, \bibinfo
  {author} {\bibfnamefont {S.}~\bibnamefont {Ghemawat}}, \bibinfo {author}
  {\bibfnamefont {I.}~\bibnamefont {Goodfellow}}, \bibinfo {author}
  {\bibfnamefont {A.}~\bibnamefont {Harp}}, \bibinfo {author} {\bibfnamefont
  {G.}~\bibnamefont {Irving}}, \bibinfo {author} {\bibfnamefont
  {M.}~\bibnamefont {Isard}}, \bibinfo {author} {\bibfnamefont
  {Y.}~\bibnamefont {Jia}}, \bibinfo {author} {\bibfnamefont {R.}~\bibnamefont
  {Jozefowicz}}, \bibinfo {author} {\bibfnamefont {L.}~\bibnamefont {Kaiser}},
  \bibinfo {author} {\bibfnamefont {M.}~\bibnamefont {Kudlur}}, \bibinfo
  {author} {\bibfnamefont {J.}~\bibnamefont {Levenberg}}, \bibinfo {author}
  {\bibfnamefont {D.}~\bibnamefont {Man\'{e}}}, \bibinfo {author}
  {\bibfnamefont {R.}~\bibnamefont {Monga}}, \bibinfo {author} {\bibfnamefont
  {S.}~\bibnamefont {Moore}}, \bibinfo {author} {\bibfnamefont
  {D.}~\bibnamefont {Murray}}, \bibinfo {author} {\bibfnamefont
  {C.}~\bibnamefont {Olah}}, \bibinfo {author} {\bibfnamefont {M.}~\bibnamefont
  {Schuster}}, \bibinfo {author} {\bibfnamefont {J.}~\bibnamefont {Shlens}},
  \bibinfo {author} {\bibfnamefont {B.}~\bibnamefont {Steiner}}, \bibinfo
  {author} {\bibfnamefont {I.}~\bibnamefont {Sutskever}}, \bibinfo {author}
  {\bibfnamefont {K.}~\bibnamefont {Talwar}}, \bibinfo {author} {\bibfnamefont
  {P.}~\bibnamefont {Tucker}}, \bibinfo {author} {\bibfnamefont
  {V.}~\bibnamefont {Vanhoucke}}, \bibinfo {author} {\bibfnamefont
  {V.}~\bibnamefont {Vasudevan}}, \bibinfo {author} {\bibfnamefont
  {F.}~\bibnamefont {Vi\'{e}gas}}, \bibinfo {author} {\bibfnamefont
  {O.}~\bibnamefont {Vinyals}}, \bibinfo {author} {\bibfnamefont
  {P.}~\bibnamefont {Warden}}, \bibinfo {author} {\bibfnamefont
  {M.}~\bibnamefont {Wattenberg}}, \bibinfo {author} {\bibfnamefont
  {M.}~\bibnamefont {Wicke}}, \bibinfo {author} {\bibfnamefont
  {Y.}~\bibnamefont {Yu}}, \ and\ \bibinfo {author} {\bibfnamefont
  {X.}~\bibnamefont {Zheng}},\ }\href {https://www.tensorflow.org/} {\enquote
  {\bibinfo {title} {{TensorFlow}: Large-scale machine learning on
  heterogeneous systems},}\ } (\bibinfo {year} {2015}),\ \bibinfo {note}
  {software available from tensorflow.org}\BibitemShut {NoStop}%
\bibitem [{\citenamefont {O'Malley}\ \emph {et~al.}(2019)\citenamefont
  {O'Malley}, \citenamefont {Bursztein}, \citenamefont {Long}, \citenamefont
  {Chollet}, \citenamefont {Jin}, \citenamefont {Invernizzi} \emph
  {et~al.}}]{omalley2019kerastuner}%
  \BibitemOpen
  \bibfield  {author} {\bibinfo {author} {\bibfnamefont {T.}~\bibnamefont
  {O'Malley}}, \bibinfo {author} {\bibfnamefont {E.}~\bibnamefont {Bursztein}},
  \bibinfo {author} {\bibfnamefont {J.}~\bibnamefont {Long}}, \bibinfo {author}
  {\bibfnamefont {F.}~\bibnamefont {Chollet}}, \bibinfo {author} {\bibfnamefont
  {H.}~\bibnamefont {Jin}}, \bibinfo {author} {\bibfnamefont {L.}~\bibnamefont
  {Invernizzi}},  \emph {et~al.},\ }\href@noop {} {\enquote {\bibinfo {title}
  {Kerastuner},}\ }\bibinfo {howpublished}
  {\url{https://github.com/keras-team/keras-tuner}} (\bibinfo {year}
  {2019})\BibitemShut {NoStop}%
\bibitem [{\citenamefont {Zhang}\ \emph {et~al.}(2015)\citenamefont {Zhang},
  \citenamefont {Wang}, \citenamefont {Zhang}, \citenamefont {Liu},
  \citenamefont {Zhong}, \citenamefont {Song}, \citenamefont {Yang},
  \citenamefont {Zhang},\ and\ \citenamefont {Ma}}]{seh3}%
  \BibitemOpen
  \bibfield  {author} {\bibinfo {author} {\bibfnamefont {S.}~\bibnamefont
  {Zhang}}, \bibinfo {author} {\bibfnamefont {Y.}~\bibnamefont {Wang}},
  \bibinfo {author} {\bibfnamefont {J.}~\bibnamefont {Zhang}}, \bibinfo
  {author} {\bibfnamefont {H.}~\bibnamefont {Liu}}, \bibinfo {author}
  {\bibfnamefont {X.}~\bibnamefont {Zhong}}, \bibinfo {author} {\bibfnamefont
  {H.-F.}\ \bibnamefont {Song}}, \bibinfo {author} {\bibfnamefont
  {G.}~\bibnamefont {Yang}}, \bibinfo {author} {\bibfnamefont {L.}~\bibnamefont
  {Zhang}}, \ and\ \bibinfo {author} {\bibfnamefont {Y.}~\bibnamefont {Ma}},\
  }\href {\doibase https://doi.org/10.1038/srep15433} {\bibfield  {journal}
  {\bibinfo  {journal} {Scientific reports}\ }\textbf {\bibinfo {volume} {5}},\
  \bibinfo {pages} {1} (\bibinfo {year} {2015})}\BibitemShut {NoStop}%
\bibitem [{\citenamefont {Cui}\ \emph {et~al.}(2020)\citenamefont {Cui},
  \citenamefont {Bi}, \citenamefont {Shi}, \citenamefont {Li}, \citenamefont
  {Liu}, \citenamefont {Zurek},\ and\ \citenamefont {Hemley}}]{Cui2020}%
  \BibitemOpen
  \bibfield  {author} {\bibinfo {author} {\bibfnamefont {W.}~\bibnamefont
  {Cui}}, \bibinfo {author} {\bibfnamefont {T.}~\bibnamefont {Bi}}, \bibinfo
  {author} {\bibfnamefont {J.}~\bibnamefont {Shi}}, \bibinfo {author}
  {\bibfnamefont {Y.}~\bibnamefont {Li}}, \bibinfo {author} {\bibfnamefont
  {H.}~\bibnamefont {Liu}}, \bibinfo {author} {\bibfnamefont {E.}~\bibnamefont
  {Zurek}}, \ and\ \bibinfo {author} {\bibfnamefont {R.~J.}\ \bibnamefont
  {Hemley}},\ }\href {\doibase 10.1103/physrevb.101.134504} {\bibfield
  {journal} {\bibinfo  {journal} {Physical Review B}\ }\textbf {\bibinfo
  {volume} {101}} (\bibinfo {year} {2020}),\
  10.1103/physrevb.101.134504}\BibitemShut {NoStop}%
\bibitem [{\citenamefont {Wang}\ \emph
  {et~al.}(2022{\natexlab{b}})\citenamefont {Wang}, \citenamefont {Wang},
  \citenamefont {Sun}, \citenamefont {Ma}, \citenamefont {Wang}, \citenamefont
  {Zou}, \citenamefont {Liu}, \citenamefont {Zhou},\ and\ \citenamefont
  {Wang}}]{Wang2022}%
  \BibitemOpen
  \bibfield  {author} {\bibinfo {author} {\bibfnamefont {Y.}~\bibnamefont
  {Wang}}, \bibinfo {author} {\bibfnamefont {K.}~\bibnamefont {Wang}}, \bibinfo
  {author} {\bibfnamefont {Y.}~\bibnamefont {Sun}}, \bibinfo {author}
  {\bibfnamefont {L.}~\bibnamefont {Ma}}, \bibinfo {author} {\bibfnamefont
  {Y.}~\bibnamefont {Wang}}, \bibinfo {author} {\bibfnamefont {B.}~\bibnamefont
  {Zou}}, \bibinfo {author} {\bibfnamefont {G.}~\bibnamefont {Liu}}, \bibinfo
  {author} {\bibfnamefont {M.}~\bibnamefont {Zhou}}, \ and\ \bibinfo {author}
  {\bibfnamefont {H.}~\bibnamefont {Wang}},\ }\href {\doibase
  10.1088/1674-1056/ac872e} {\bibfield  {journal} {\bibinfo  {journal} {Chinese
  Physics B}\ }\textbf {\bibinfo {volume} {31}},\ \bibinfo {pages} {106201}
  (\bibinfo {year} {2022}{\natexlab{b}})}\BibitemShut {NoStop}%
\bibitem [{\citenamefont {Borinaga}\ \emph {et~al.}(2016)\citenamefont
  {Borinaga}, \citenamefont {Errea}, \citenamefont {Calandra}, \citenamefont
  {Mauri},\ and\ \citenamefont {Bergara}}]{anharmonich3}%
  \BibitemOpen
  \bibfield  {author} {\bibinfo {author} {\bibfnamefont {M.}~\bibnamefont
  {Borinaga}}, \bibinfo {author} {\bibfnamefont {I.}~\bibnamefont {Errea}},
  \bibinfo {author} {\bibfnamefont {M.}~\bibnamefont {Calandra}}, \bibinfo
  {author} {\bibfnamefont {F.}~\bibnamefont {Mauri}}, \ and\ \bibinfo {author}
  {\bibfnamefont {A.}~\bibnamefont {Bergara}},\ }\href {\doibase
  10.1103/PhysRevB.93.174308} {\bibfield  {journal} {\bibinfo  {journal} {Phys.
  Rev. B}\ }\textbf {\bibinfo {volume} {93}},\ \bibinfo {pages} {174308}
  (\bibinfo {year} {2016})}\BibitemShut {NoStop}%
\bibitem [{Note1()}]{Note1}%
  \BibitemOpen
  \bibinfo {note} {Structure files are available by \protect \url
  {https://github.com/LazarNov/superconductor-predict}. Atoms labelled as X/Y
  specify the morphed sites.}\BibitemShut {Stop}%
\bibitem [{\citenamefont {Abe}\ and\ \citenamefont {Ashcroft}(2011)}]{BH3}%
  \BibitemOpen
  \bibfield  {author} {\bibinfo {author} {\bibfnamefont {K.}~\bibnamefont
  {Abe}}\ and\ \bibinfo {author} {\bibfnamefont {N.~W.}\ \bibnamefont
  {Ashcroft}},\ }\href {\doibase 10.1103/PhysRevB.84.104118} {\bibfield
  {journal} {\bibinfo  {journal} {Phys. Rev. B}\ }\textbf {\bibinfo {volume}
  {84}},\ \bibinfo {pages} {104118} (\bibinfo {year} {2011})}\BibitemShut
  {NoStop}%
\bibitem [{\citenamefont {Drozdov}\ \emph
  {et~al.}(2015{\natexlab{b}})\citenamefont {Drozdov}, \citenamefont
  {Eremets},\ and\ \citenamefont {Troyan}}]{ph}%
  \BibitemOpen
  \bibfield  {author} {\bibinfo {author} {\bibfnamefont {A.~P.}\ \bibnamefont
  {Drozdov}}, \bibinfo {author} {\bibfnamefont {M.~I.}\ \bibnamefont
  {Eremets}}, \ and\ \bibinfo {author} {\bibfnamefont {I.~A.}\ \bibnamefont
  {Troyan}},\ }\href {\doibase 10.48550/ARXIV.1508.06224} {\enquote {\bibinfo
  {title} {Superconductivity above 100 k in ph3 at high pressures},}\ }
  (\bibinfo {year} {2015}{\natexlab{b}})\BibitemShut {NoStop}%
\bibitem [{\citenamefont {Giannozzi}\ \emph {et~al.}(2009)\citenamefont
  {Giannozzi}, \citenamefont {Baroni}, \citenamefont {Bonini}, \citenamefont
  {Calandra}, \citenamefont {Car}, \citenamefont {Cavazzoni}, \citenamefont
  {Ceresoli}, \citenamefont {Chiarotti}, \citenamefont {Cococcioni},
  \citenamefont {Dabo}, \citenamefont {Corso}, \citenamefont {de~Gironcoli},
  \citenamefont {Fabris}, \citenamefont {Fratesi}, \citenamefont {Gebauer},
  \citenamefont {Gerstmann}, \citenamefont {Gougoussis}, \citenamefont
  {Kokalj}, \citenamefont {Lazzeri}, \citenamefont {Martin-Samos},
  \citenamefont {Marzari}, \citenamefont {Mauri}, \citenamefont {Mazzarello},
  \citenamefont {Paolini}, \citenamefont {Pasquarello}, \citenamefont
  {Paulatto}, \citenamefont {Sbraccia}, \citenamefont {Scandolo}, \citenamefont
  {Sclauzero}, \citenamefont {Seitsonen}, \citenamefont {Smogunov},
  \citenamefont {Umari},\ and\ \citenamefont {Wentzcovitch}}]{Giannozzi_2009}%
  \BibitemOpen
  \bibfield  {author} {\bibinfo {author} {\bibfnamefont {P.}~\bibnamefont
  {Giannozzi}}, \bibinfo {author} {\bibfnamefont {S.}~\bibnamefont {Baroni}},
  \bibinfo {author} {\bibfnamefont {N.}~\bibnamefont {Bonini}}, \bibinfo
  {author} {\bibfnamefont {M.}~\bibnamefont {Calandra}}, \bibinfo {author}
  {\bibfnamefont {R.}~\bibnamefont {Car}}, \bibinfo {author} {\bibfnamefont
  {C.}~\bibnamefont {Cavazzoni}}, \bibinfo {author} {\bibfnamefont
  {D.}~\bibnamefont {Ceresoli}}, \bibinfo {author} {\bibfnamefont {G.~L.}\
  \bibnamefont {Chiarotti}}, \bibinfo {author} {\bibfnamefont {M.}~\bibnamefont
  {Cococcioni}}, \bibinfo {author} {\bibfnamefont {I.}~\bibnamefont {Dabo}},
  \bibinfo {author} {\bibfnamefont {A.~D.}\ \bibnamefont {Corso}}, \bibinfo
  {author} {\bibfnamefont {S.}~\bibnamefont {de~Gironcoli}}, \bibinfo {author}
  {\bibfnamefont {S.}~\bibnamefont {Fabris}}, \bibinfo {author} {\bibfnamefont
  {G.}~\bibnamefont {Fratesi}}, \bibinfo {author} {\bibfnamefont
  {R.}~\bibnamefont {Gebauer}}, \bibinfo {author} {\bibfnamefont
  {U.}~\bibnamefont {Gerstmann}}, \bibinfo {author} {\bibfnamefont
  {C.}~\bibnamefont {Gougoussis}}, \bibinfo {author} {\bibfnamefont
  {A.}~\bibnamefont {Kokalj}}, \bibinfo {author} {\bibfnamefont
  {M.}~\bibnamefont {Lazzeri}}, \bibinfo {author} {\bibfnamefont
  {L.}~\bibnamefont {Martin-Samos}}, \bibinfo {author} {\bibfnamefont
  {N.}~\bibnamefont {Marzari}}, \bibinfo {author} {\bibfnamefont
  {F.}~\bibnamefont {Mauri}}, \bibinfo {author} {\bibfnamefont
  {R.}~\bibnamefont {Mazzarello}}, \bibinfo {author} {\bibfnamefont
  {S.}~\bibnamefont {Paolini}}, \bibinfo {author} {\bibfnamefont
  {A.}~\bibnamefont {Pasquarello}}, \bibinfo {author} {\bibfnamefont
  {L.}~\bibnamefont {Paulatto}}, \bibinfo {author} {\bibfnamefont
  {C.}~\bibnamefont {Sbraccia}}, \bibinfo {author} {\bibfnamefont
  {S.}~\bibnamefont {Scandolo}}, \bibinfo {author} {\bibfnamefont
  {G.}~\bibnamefont {Sclauzero}}, \bibinfo {author} {\bibfnamefont {A.~P.}\
  \bibnamefont {Seitsonen}}, \bibinfo {author} {\bibfnamefont {A.}~\bibnamefont
  {Smogunov}}, \bibinfo {author} {\bibfnamefont {P.}~\bibnamefont {Umari}}, \
  and\ \bibinfo {author} {\bibfnamefont {R.~M.}\ \bibnamefont {Wentzcovitch}},\
  }\href {\doibase 10.1088/0953-8984/21/39/395502} {\bibfield  {journal}
  {\bibinfo  {journal} {J. Phys.: Cond. Matter}\ }\textbf {\bibinfo {volume}
  {21}},\ \bibinfo {pages} {395502} (\bibinfo {year} {2009})}\BibitemShut
  {NoStop}%
\bibitem [{\citenamefont {Giannozzi}\ \emph {et~al.}(2017)\citenamefont
  {Giannozzi}, \citenamefont {Andreussi}, \citenamefont {Brumme}, \citenamefont
  {Bunau}, \citenamefont {Nardelli}, \citenamefont {Calandra}, \citenamefont
  {Car}, \citenamefont {Cavazzoni}, \citenamefont {Ceresoli}, \citenamefont
  {Cococcioni}, \citenamefont {Colonna}, \citenamefont {Carnimeo},
  \citenamefont {Corso}, \citenamefont {de~Gironcoli}, \citenamefont {Delugas},
  \citenamefont {DiStasio}, \citenamefont {Ferretti}, \citenamefont {Floris},
  \citenamefont {Fratesi}, \citenamefont {Fugallo}, \citenamefont {Gebauer},
  \citenamefont {Gerstmann}, \citenamefont {Giustino}, \citenamefont {Gorni},
  \citenamefont {Jia}, \citenamefont {Kawamura}, \citenamefont {Ko},
  \citenamefont {Kokalj}, \citenamefont {Kü{\c{c}}ükbenli}, \citenamefont
  {Lazzeri}, \citenamefont {Marsili}, \citenamefont {Marzari}, \citenamefont
  {Mauri}, \citenamefont {Nguyen}, \citenamefont {Nguyen}, \citenamefont
  {de-la-Roza~andd L~Paulatto}, \citenamefont {Ponc{\'{e}}}, \citenamefont
  {Rocca}, \citenamefont {Sabatini}, \citenamefont {Santra}, \citenamefont
  {Schlipf}, \citenamefont {Seitsonen}, \citenamefont {Smogunov}, \citenamefont
  {Timrov}, \citenamefont {Thonhauser}, \citenamefont {Umari}, \citenamefont
  {Vast}, \citenamefont {Wu},\ and\ \citenamefont {Baroni}}]{Giannozzi_2017}%
  \BibitemOpen
  \bibfield  {author} {\bibinfo {author} {\bibfnamefont {P.}~\bibnamefont
  {Giannozzi}}, \bibinfo {author} {\bibfnamefont {O.}~\bibnamefont
  {Andreussi}}, \bibinfo {author} {\bibfnamefont {T.}~\bibnamefont {Brumme}},
  \bibinfo {author} {\bibfnamefont {O.}~\bibnamefont {Bunau}}, \bibinfo
  {author} {\bibfnamefont {M.~B.}\ \bibnamefont {Nardelli}}, \bibinfo {author}
  {\bibfnamefont {M.}~\bibnamefont {Calandra}}, \bibinfo {author}
  {\bibfnamefont {R.}~\bibnamefont {Car}}, \bibinfo {author} {\bibfnamefont
  {C.}~\bibnamefont {Cavazzoni}}, \bibinfo {author} {\bibfnamefont
  {D.}~\bibnamefont {Ceresoli}}, \bibinfo {author} {\bibfnamefont
  {M.}~\bibnamefont {Cococcioni}}, \bibinfo {author} {\bibfnamefont
  {N.}~\bibnamefont {Colonna}}, \bibinfo {author} {\bibfnamefont
  {I.}~\bibnamefont {Carnimeo}}, \bibinfo {author} {\bibfnamefont {A.~D.}\
  \bibnamefont {Corso}}, \bibinfo {author} {\bibfnamefont {S.}~\bibnamefont
  {de~Gironcoli}}, \bibinfo {author} {\bibfnamefont {P.}~\bibnamefont
  {Delugas}}, \bibinfo {author} {\bibfnamefont {R.~A.}\ \bibnamefont
  {DiStasio}}, \bibinfo {author} {\bibfnamefont {A.}~\bibnamefont {Ferretti}},
  \bibinfo {author} {\bibfnamefont {A.}~\bibnamefont {Floris}}, \bibinfo
  {author} {\bibfnamefont {G.}~\bibnamefont {Fratesi}}, \bibinfo {author}
  {\bibfnamefont {G.}~\bibnamefont {Fugallo}}, \bibinfo {author} {\bibfnamefont
  {R.}~\bibnamefont {Gebauer}}, \bibinfo {author} {\bibfnamefont
  {U.}~\bibnamefont {Gerstmann}}, \bibinfo {author} {\bibfnamefont
  {F.}~\bibnamefont {Giustino}}, \bibinfo {author} {\bibfnamefont
  {T.}~\bibnamefont {Gorni}}, \bibinfo {author} {\bibfnamefont
  {J.}~\bibnamefont {Jia}}, \bibinfo {author} {\bibfnamefont {M.}~\bibnamefont
  {Kawamura}}, \bibinfo {author} {\bibfnamefont {H.-Y.}\ \bibnamefont {Ko}},
  \bibinfo {author} {\bibfnamefont {A.}~\bibnamefont {Kokalj}}, \bibinfo
  {author} {\bibfnamefont {E.}~\bibnamefont {Kü{\c{c}}ükbenli}}, \bibinfo
  {author} {\bibfnamefont {M.}~\bibnamefont {Lazzeri}}, \bibinfo {author}
  {\bibfnamefont {M.}~\bibnamefont {Marsili}}, \bibinfo {author} {\bibfnamefont
  {N.}~\bibnamefont {Marzari}}, \bibinfo {author} {\bibfnamefont
  {F.}~\bibnamefont {Mauri}}, \bibinfo {author} {\bibfnamefont {N.~L.}\
  \bibnamefont {Nguyen}}, \bibinfo {author} {\bibfnamefont {H.-V.}\
  \bibnamefont {Nguyen}}, \bibinfo {author} {\bibfnamefont {A.~O.}\
  \bibnamefont {de-la-Roza~andd L~Paulatto}}, \bibinfo {author} {\bibfnamefont
  {S.}~\bibnamefont {Ponc{\'{e}}}}, \bibinfo {author} {\bibfnamefont
  {D.}~\bibnamefont {Rocca}}, \bibinfo {author} {\bibfnamefont
  {R.}~\bibnamefont {Sabatini}}, \bibinfo {author} {\bibfnamefont
  {B.}~\bibnamefont {Santra}}, \bibinfo {author} {\bibfnamefont
  {M.}~\bibnamefont {Schlipf}}, \bibinfo {author} {\bibfnamefont {A.~P.}\
  \bibnamefont {Seitsonen}}, \bibinfo {author} {\bibfnamefont {A.}~\bibnamefont
  {Smogunov}}, \bibinfo {author} {\bibfnamefont {I.}~\bibnamefont {Timrov}},
  \bibinfo {author} {\bibfnamefont {T.}~\bibnamefont {Thonhauser}}, \bibinfo
  {author} {\bibfnamefont {P.}~\bibnamefont {Umari}}, \bibinfo {author}
  {\bibfnamefont {N.}~\bibnamefont {Vast}}, \bibinfo {author} {\bibfnamefont
  {X.}~\bibnamefont {Wu}}, \ and\ \bibinfo {author} {\bibfnamefont
  {S.}~\bibnamefont {Baroni}},\ }\href {\doibase 10.1088/1361-648x/aa8f79}
  {\bibfield  {journal} {\bibinfo  {journal} {J. Phys.: Cond. Matter}\ }\textbf
  {\bibinfo {volume} {29}},\ \bibinfo {pages} {465901} (\bibinfo {year}
  {2017})}\BibitemShut {NoStop}%
\bibitem [{\citenamefont {Zhu}\ \emph {et~al.}(2022)\citenamefont {Zhu},
  \citenamefont {Yang}, \citenamefont {Xia}, \citenamefont {Yin}, \citenamefont
  {Wang}, \citenamefont {Zhao}, \citenamefont {Dai}, \citenamefont {Tu},
  \citenamefont {Song}, \citenamefont {Tao}, \citenamefont {Tu}, \citenamefont
  {Gong}, \citenamefont {Lei}, \citenamefont {Guo},\ and\ \citenamefont
  {Li}}]{PhysRevB.105.094507}%
  \BibitemOpen
  \bibfield  {author} {\bibinfo {author} {\bibfnamefont {C.~C.}\ \bibnamefont
  {Zhu}}, \bibinfo {author} {\bibfnamefont {X.~F.}\ \bibnamefont {Yang}},
  \bibinfo {author} {\bibfnamefont {W.}~\bibnamefont {Xia}}, \bibinfo {author}
  {\bibfnamefont {Q.~W.}\ \bibnamefont {Yin}}, \bibinfo {author} {\bibfnamefont
  {L.~S.}\ \bibnamefont {Wang}}, \bibinfo {author} {\bibfnamefont {C.~C.}\
  \bibnamefont {Zhao}}, \bibinfo {author} {\bibfnamefont {D.~Z.}\ \bibnamefont
  {Dai}}, \bibinfo {author} {\bibfnamefont {C.~P.}\ \bibnamefont {Tu}},
  \bibinfo {author} {\bibfnamefont {B.~Q.}\ \bibnamefont {Song}}, \bibinfo
  {author} {\bibfnamefont {Z.~C.}\ \bibnamefont {Tao}}, \bibinfo {author}
  {\bibfnamefont {Z.~J.}\ \bibnamefont {Tu}}, \bibinfo {author} {\bibfnamefont
  {C.~S.}\ \bibnamefont {Gong}}, \bibinfo {author} {\bibfnamefont {H.~C.}\
  \bibnamefont {Lei}}, \bibinfo {author} {\bibfnamefont {Y.~F.}\ \bibnamefont
  {Guo}}, \ and\ \bibinfo {author} {\bibfnamefont {S.~Y.}\ \bibnamefont {Li}},\
  }\href {\doibase 10.1103/PhysRevB.105.094507} {\bibfield  {journal} {\bibinfo
   {journal} {Phys. Rev. B}\ }\textbf {\bibinfo {volume} {105}},\ \bibinfo
  {pages} {094507} (\bibinfo {year} {2022})}\BibitemShut {NoStop}%
\end{thebibliography}%
\end{document}